\documentclass[lettersize,journal]{IEEEtran}
\usepackage{amsmath,amssymb,amsfonts}
\usepackage{graphicx}
\usepackage{textcomp}
\usepackage[colorlinks,urlcolor=blue,linkcolor=blue,citecolor=blue]{hyperref}
\usepackage{cite}
\usepackage{color,array}
\usepackage{amsmath}
\usepackage{algorithm}
\usepackage{stfloats}
\usepackage{url}
\usepackage{verbatim}
\usepackage{graphicx}
\usepackage{algpseudocode}
\usepackage{subcaption}
\usepackage{amsmath, amssymb, amsfonts}
\usepackage{graphicx}
\usepackage{multicol}
\usepackage{multirow}
\usepackage{makecell}
\usepackage{balance}
\usepackage[section]{placeins}
\usepackage{tipa}
\usepackage{xcolor}
\usepackage{svg}
\usepackage{pdfpages}
\usepackage{tabularx}
\usepackage{hyperref}
\usepackage{booktabs}
\usepackage{nicefrac}
\usepackage{microtype}
\usepackage{wrapfig}
\usepackage[normalem]{ulem}
\usepackage{colortbl}
\usepackage{newunicodechar}
\usepackage{changepage}
\usepackage{xcolor}
\usepackage{amsmath}
\providecommand{\orgdiv}[1]{#1}
\providecommand{\orgname}[1]{#1}

\DeclareMathOperator*{\argmax}{arg\,max}

\begin{document}

\title{Parallel Delayed Memory Units for Enhanced Temporal Modeling in Biomedical and Bioacoustic Signal Analysis} 

\author{Pengfei Sun, Wenyu Jiang, \IEEEmembership{Senior Member, IEEE}, Paul Devos, and Dick Botteldooren, \IEEEmembership{Senior Member, IEEE}
\thanks{This work was supported in part by the Flemish Government under the "Onderzoeksprogramma Artificiele Intelligentie (AI) Vlaanderen" and the Research Foundation - Flanders under grant number G0A0220N, and in part by Programmatic grant no. A1687b0033 from the Singapore government's Research, Innovation and Enterprise 2020 plan (Advanced Manufacturing and Engineering domain). Corresponding author:  dick.botteldooren@ugent.be.}
\thanks{Pengfei Sun, Paul Devos, and Dick Botteldooren are with the Department of Information Technology, WAVES Research Group, Ghent University. Present address (Pengfei Sun): {\orgdiv{Department of Electrical and Electronic Engineering}, \orgname{Imperial College London}} (e-mail: p.sun@imperial.ac.uk).}
\thanks{Wenyu Jiang is with Institute for Infocomm Research (I$^2$R), Agency for Science, Technology and Research (A*STAR), Singapore}
}

\maketitle

\begin{abstract}
Advanced deep learning architectures, particularly recurrent neural networks (RNNs), have been widely applied in audio, bioacoustic, and biomedical signal analysis, especially in data-scarce environments. While gated RNNs remain effective, they can be relatively over-parameterised and less training-efficient in some regimes \cite{ravanelli2017improving,sun2023delayed}, while linear RNNs tend to fall short in capturing the complexity inherent in bio-signals. To address these challenges, we propose the Parallel Delayed Memory Unit (PDMU), a {delay-gated state-space module for short-term temporal credit assignment} targeting audio and bioacoustic signals, which enhances short-term temporal state interactions and memory efficiency via a gated delay-line mechanism. Unlike previous Delayed Memory Units (DMU) that embed temporal dynamics into the delay-line architecture, the PDMU further compresses temporal information into vector representations using Legendre Memory Units (LMU). This design serves as a form of causal attention, allowing the model to dynamically adjust its reliance on past states and improve real-time learning performance. Notably, in low-information scenarios, the gating mechanism behaves similarly to skip connections by bypassing state decay and preserving early representations, thereby facilitating long-term memory retention. The PDMU is modular, supporting parallel training and sequential inference, and can be easily integrated into existing linear RNN frameworks. Furthermore, we introduce bidirectional, efficient, and spiking variants of the architecture, each offering additional gains in performance or energy efficiency. Experimental results on diverse audio and biomedical benchmarks demonstrate that the PDMU significantly enhances both memory capacity and overall model performance.

\end{abstract}


\section{Introduction}
\IEEEPARstart{A}{udio}, bioacoustic, and biomedical signals contain valuable latent information that is essential for understanding and monitoring individual health conditions \cite{fernando2022deep, qian2024learning}. Recent advances in machine learning have greatly improved the processing and analysis of audio and biomedical signals~\cite{fernando2022deep, qian2024learning,vaiciukynas2017detecting,de2022no}. Gated RNNs \cite{hochreiter1997long}, designed specifically for sequential data, offer a low-complexity solution with a time complexity of O(N). Gated RNNs process input signals by feeding them into memory cells that aggregate and redistribute information over time through complex network dynamics. Another widely used deep learning model is the Transformer \cite{vaswani2017attention,9844844}, which employs a self-attention mechanism to assign importance to different input elements, effectively capturing long-term dependencies in the data. Unlike RNNs, Transformers are stateless architectures, allowing for parallel training and significantly accelerating the training process. RNNs and Transformers are particularly well-suited for audio-based tasks, as they
can handle the temporal dependencies that are critical for sequential data processing. These architectures, along with their various adaptations, form the backbone of many audio-based deep learning models. 

Despite considerable advancements in applying recurrent neural networks to bio-signal tasks, several challenges remain. The increasing volume of data necessitates faster processing capabilities, as well as the ability to effectively address tasks that demand the integration of both long-term and short-term temporal dependencies. Moreover, for these models to be practical in real-world environments, they should be optimized for deployment in resource-limited devices and edge computing scenarios, where lightweight architectures and real-time inference are essential. However, current models often fail to fully meet these requirements of training speed and efficiency needed in real-world healthcare environments. Gated RNNs, while effective in various contexts, are hindered by the vanishing gradient problem, which compromises their ability to capture and maintain long-term dependencies across extended sequences. The nonlinear memory state updates inherent in Gated RNNs also result in longer training times and pose significant challenges for parallelization, limiting their overall efficiency. Transformer-based architectures, although proficient at modeling long-term dependencies through self-attention mechanisms, require the long sequence to be available before computation can begin. This constraint not only impairs their ability to process sequential data efficiently but also leads to computational and memory overheads that scale quadratically with sequence length,  making them less suitable for energy-constrained environments \cite{zheng2020improving}. Carefully designed state-space models (e.g., Mamba \cite{gu2023mamba}) can be competitive. Legendre Memory Units (LMUs) \cite{voelker2019legendre, chilkuri2021parallelizing}, a type of linear RNN, have emerged as a promising solution for retaining memory information over theoretically infinite time horizons. They achieve this by integrating a linear time-invariant (LTI) system, thereby enabling efficient parallelization while maintaining the sequential nature of inference. However, despite their strength in handling very long-term dependencies, this kind of state-space model does not fully leverage the dynamic, short-term interactions that are vital for processing audio and biomedical signals \cite{10423155}. In light of these limitations, this work aims to enhance the functionality of linear RNNs for bio-signal processing by increasing their memory capacity for short-term memory interactions.

In this paper, we propose a novel RNN variant referred to as the Parallel Delayed Memory Unit (PDMU). This architecture builds upon the Legendre Memory Unit. The PDMU supports parallel training while processing real-time data sequentially. Additionally, the delayed memory unit facilitates interaction between the hidden states in the temporal dimension and enhances temporal credit assignment, thereby improving the memory capacity of the RNN. To further improve performance and reduce computational complexity, several variants of the PDMU are proposed and analyzed. These modifications achieve state-of-the-art or competitive performance compared to other works and show significant improvement over the baseline. Our primary contributions can be summarized as follows:

\begin{itemize}
\item We introduce a novel Parallel Delayed Memory Unit (PDMU), a delay-gated state-space module that explicitly captures short-horizon temporal interactions. Coupled with a linear RNN cell, PDMU retains fully parallel training while preserving causal, real-time inference.

\item We devised several variants to further enhance performance or improve energy efficiency, including:
\begin{itemize}
\item Bi-directional PDMU (Bi-PDMU): Captures temporal relationships from both the past and the future. Choose for offline/bidirectional settings.
\item Efficient PDMU (EPDMU): Activates only one delay gate at a time to optimize energy usage. Choose when the compute/energy budget is tight (lower MACs).
\item Spiking DMU: Incorporates spiking neurons to further increase energy efficiency. Choose for neuromorphic/low-power deployment or event-based inputs.
\end{itemize}

\item We demonstrate PDMU's superior performance across a range of audio and bio-signal benchmark tasks, particularly in scenarios with limited data, as commonly encountered in real-world healthcare environments where optimization for edge computing deployment is crucial. PDMU not only significantly enhances signal decoding but also consistently outperforms traditional RNNs and other baseline models. 
\end{itemize}

The structure of this paper is as follows: Section \ref{related} provides the relevant methods, situating our work within the current research context. Section \ref{method} details the proposed methodology. Section \ref{experiment} presents the evaluation of our approach on five benchmark tasks, including an ablation study and model analysis. Finally, Section \ref{discussion} offers a summary and conclusions of the study.

\section{Related work}
\label{related}
\subsection{Legendre Memory Unit and Parallel Training}
The Legendre Memory Unit (LMU), introduced by Voelker et al. \cite{voelker2019legendre}, effectively captures and encodes temporal dependencies in sequential data through the use of Legendre polynomials. This memory cell is constructed by integrating a single-input delay network (DN) with a nonlinear dynamical system. The DN orthogonalizes the input signal over a sliding window of length $\theta$, while the nonlinear system leverages this orthogonalized memory to compute various functions over time. Mathematically, given an input scalar function {u}(t), the state update can be described as follows:
\begin{equation}
\mathbf{m}'(t) = \mathbf{A}\mathbf{m}(t) + \mathbf{B}{u}(t)
\end{equation}
where $\textbf{m}(t) \in R^d$ denotes the memory state vector with dimension $d$, and A and B are state-space matrices. Following the use of Pad{\'e} approximation \cite{pade1892representation}, the state-space matrices can be expressed as follows: 

\begin{equation}
\label{pade}
\begin{aligned}
\mathbf{A} &= \left[ a \right]_{ij} \in \mathbb{R}^{d \times d}, \quad a_{ij} = (2i + 1) \left\{ 
\begin{array}{ll}
-1 & \text{if } i < j \\
(-1)^{i-j+1} & \text{if } i \geq j 
\end{array} \right. \\
\mathbf{B} &= \left[ b \right]_{i} \in \mathbb{R}^{d \times 1}, \quad b_{i} = (2i + 1)(-1)^{i}, \quad i,j \in [0, d-1]
\end{aligned}
\end{equation}
This continuous-time system can be converted to discrete-time $k$ with a time resolution $\Delta t$:
\begin{equation}
\mathbf{m}[k] = \mathbf{\bar{A}} \mathbf{m}[k-1] + \mathbf{\bar{B}} {u}[k] 
\end{equation}
Here, $\bar{{\textbf{A}}}$ and $\bar{{\textbf{B}}}$ are the discretized versions \footnote{Using zero-order hold (ZOH) method, exact discretization gives
\begin{equation}
\bar{\textbf{A}} = e^{\textbf{A} \Delta t} \quad \text{and} \quad \bar{\textbf{B}} = \textbf{A}^{-1} (e^{\textbf{A} \Delta t} - \textbf{I}) \textbf{B}.
\end{equation}
} of $\textbf{A}$ and $\textbf{B}$, which are usually frozen during training. As suggested by Chilkuri et al. \cite{chilkuri2021parallelizing}, to enhance the efficiency of parallel training, we adopt the strategy of inputting the signal $x[k]$ into the LMU cell as follows: 
\begin{equation}
\label{input}
    {u}[k] = {f_u(\mathbf{W_{u}}\mathbf{x}[k]+\mathbf{b_{u}})}
\end{equation}
The output of this cell is described as:
\begin{equation}
    \mathbf{o}[k] = {f_o(\mathbf{W_{x}}\textbf{x}[k]+\mathbf{W_{m}}\textbf{m}[k]+\mathbf{b_o})}
\end{equation}
where \( W_{u} \), \( W_{x} \), and \( W_{m} \) are trainable weights, $b_o$ and $b_u$ are the trainable bias, and \( f_u \) and \( f_o \) are activation functions. Since Eqs. (1-3)  are linear and time-invariant (LTI), the state update equation can be expressed in a non-sequential fashion, which is crucial for enabling parallel training. Additionally, the fast Fourier transform (FFT) is employed to further reduce training complexity \cite{chilkuri2021parallelizing}.

\subsection{Spiking Neural Networks}
Spiking Neural Networks (SNNs) represent the third generation of neural networks \cite{maass1997networks}, and have garnered increasing attention due to their sparse activation patterns and the development of specialized hardware \cite{shrestha2024efficient}. In contrast to traditional artificial neurons, spiking neurons accumulate input spike trains, which maintain the membrane potential, functioning as a form of temporal memory. When the membrane potential surpasses a certain threshold, the neuron emits a spike that is propagated to the subsequent layer, followed by a reset of the potential to a predefined baseline. This dynamic enables spiking neurons to model spatial-temporal relationships effectively \cite{wang2023spatial}, while also offering greater energy efficiency relative to artificial neurons \cite{shrestha2024efficient}. Large spiking networks are typically trained with surrogate gradient descent~\cite{neftci2019surrogate}, which has been shown to exploit  spike timing information~\cite{yu2025beyond}.

Historically, SNNs have predominantly been built upon recurrent neural network architectures \cite{yin2021accurate}. Enhancements to these models have been realized through the incorporation of synaptic and axonal delays \cite{shrestha2018slayer, zhang2020supervised,sun2022axonal,sun2025exploitingheterogeneousdelaysefficient}, heterogeneous time constant \cite{perez2021neural}, as well as adaptive delays and thresholds \cite{yin2021accurate,10094768}, which contribute to their flexibility and performance. Furthermore, spiking neurons are inherently capable of processing spike-based information directly. However, when addressing data that contains multi-bit information, it is necessary to introduce an encoding layer that converts these inputs into spike trains \cite{wu2021tandem}. In this work, for tasks involving multi-bit inputs, we utilize an encoding layer that transforms these inputs into spike trains.

Recently, there has been a growing interest in using SNNs for sequential learning tasks. As more advanced architectures have emerged, researchers have begun exploring the integration of spiking neurons within structured state-space models. For example, Du et al. \cite{du2024spiking} have successfully combined these models with SNNs, demonstrating significant potential to enhance both performance and efficiency. Nevertheless, the optimal methods for incorporating spiking neurons into state-space models remain an open area of research.  

\begin{figure*}[t]

    \centering
 \includegraphics[width=0.58\linewidth]{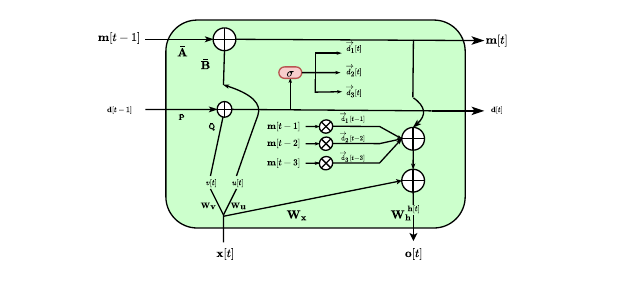}
  \includegraphics[width=0.40\linewidth]{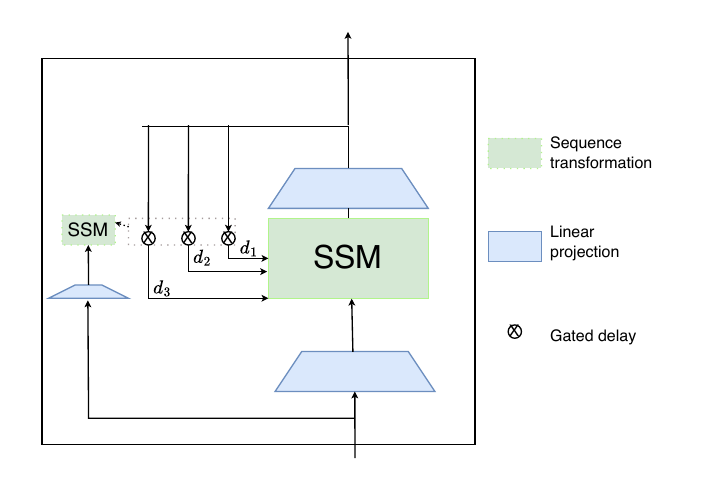}
    \caption{{{Time-unrolled PDMU Cell (\textbf{left}) and simplified block diagram(\textbf{right}): A d-dimensional state vector ($h[t]$) is dynamically coupled with a d-dimensional memory vector ($m[t]$) and a delayed memory vector ($\overset{\rightarrow}d_{t-i}[i]m[t-i]$). At each step, the gate produces $n$ coefficients (here $n{=}3$) that route the current memory to specific future offsets—i.e., learnable delays.}}}
    
    \label{fig:gate}
    
\end{figure*}

\section{Method}
\label{method}
\subsection{Parallel Delayed Memory Unit}

The Parallel Delayed Memory Unit (PDMU), illustrated in Figure \ref{fig:gate}, is designed to facilitate parallel training while supporting sequential data processing during inference. Building upon the Legendre Memory Unit (LMU) framework, our model incorporates a delay line architecture equipped with a gating function that distributes current memory information to optimal future time points, enhancing the network's capacity to capture temporal dependencies.

The core component of PDMU is the delay gate, which controls the flow of memory information along the delay lines. To support parallel training, we employ a lightweight state-space model to achieve this:
\begin{equation}
\mathbf{d}'(t) = \mathbf{P}\mathbf{d}(t) + \mathbf{Q}{{v}}(t)
\end{equation}
where $\textbf{d}(t) \in R^n$ denotes the delay memory state vector with a length of $n$, and \textbf{P} and \textbf{Q} are state-space matrices similar to \textbf{A} and \textbf{B} except for their dimensions. Following the previous discretization, it can be expressed as: 
\begin{equation}
\label{cell_delay}
    \textbf{{d}}[k] = \bar{{\textbf{P}}} \textbf{d}[k-1] + \bar{{\textbf{Q}}} {{v}}[k] 
\end{equation}
where ${{v}}[k] = f_u(\textbf{W}_\textbf{v}\textbf{x}[k]+\textbf{b}_\textbf{v})$, with $\textbf{W}_\textbf{v}$ and $\textbf{b}_\textbf{v}$ denoting the trainable weight matrix and bias vector, respectively. The output memory state $\textbf{h}[k]$ is obtained not only from the long-term memory $\textbf{m}[k]$ but also from the short-term delayed memory. This can be expressed as: 

\begin{equation}
\label{pdmu_1}
\mathbf{h}[k] = \mathbf{m}[k]
+ \sum_{i=k-n}^{k-1} \overset{\rightarrow}{{d}}_{k-i}[i] \,\mathbf{m}[i]
\end{equation}

Here, we denote '\(\rightarrow\)' as the information flow direction for the future. Then, the output of the PDMU module can be described as:

\begin{equation}
\label{pmdu_5}
    \mathbf{o}[k] = f_o(\mathbf{W_{h}}\textbf{h}[k]+\mathbf{W_{x}}\textbf{x}[k]+\mathbf{b_{o}})
\end{equation}
For the parallel training,  because this is an LTI
system, standard control theory gives us a non-iterative way of evaluating this equation \cite{chilkuri2021parallelizing} as shown below:   
\begin{equation}
\label{memeoryclee}
\textbf{{m}}[k] = \sum_{j=1}^{k} {\bar{\textbf{A}}}^{k-j} {\bar{\textbf{B}} {u}_j}
\end{equation}
and
\begin{equation}
\label{delay_cell}
\textbf{{d}}[k] = {\sum_{j=1}^{k} {\bar{\textbf{P}}}^{k-j} {\bar{\textbf{Q}} \bar {{v}}_j}}
\end{equation}
Here, we define the delay gate as \(\overset{\rightarrow}{{d}}_{i}[k] = \sigma_{i}(\textbf{d}[k])\), where \(i \in \{1, \dots, n\}\), and \(\sigma_{i}(\cdot)\) denotes its \(i\)-th output component (e.g., Softmax).
 For better training and visualization,  the output hidden state of the PDMU module can alternatively be written as a matrix multiplication and summation. Assuming $\textbf{H} = [H[1], H[2], \cdots, H[T]]$, each $H[k]$ is the result of the element-wise multiplication and summation of the $k^{\textrm{th}}$ columns of matrices $\hat{\textbf{D}}$ and $\hat{\textbf{M}}$. 
\begin{equation}
\mathbf{H}[k] = \sum_{i=k-n}^{k} \mathbf{\hat{D}}_{i,k} \circ \mathbf{\hat{M}}_{i,k}
\end{equation}
where

\begin{equation}
\mathbf{\hat{D}} = \left[ \begin{array}{*{8}{c@{\hspace{0.03em}}}}
1 & \overset{\rightarrow}d_{1}[1] & \overset{\rightarrow}d_{2}[1] & \cdots & \overset{\rightarrow}d_{n}[1] & \cdots  & 0 & 0 \\
0 & 1 &\overset{\rightarrow} d_{1}[2] & \cdots & \overset{\rightarrow}d_{n-1}[2] & \cdots  & 0 & 0 \\
0 & 0 & 1 & \cdots & \overset{\rightarrow}d_{n-2}[3]  & \cdots  & 0 & 0 \\
\vdots & \vdots & \vdots & \ddots & \vdots &  \vdots & \vdots & \vdots \\
0 & 0 & 0 & \cdots & 0  & \cdots &\overset{\rightarrow} d_{1}[T-2]  &\overset{\rightarrow} d_{2}[T-2] \\
0 & 0 & 0 & \cdots & 0 & \cdots  & 1 & \overset{\rightarrow}d_{1}[T-1] \\
0 & 0 & 0 & \cdots & 0 & \cdots  & 0 & 1 
\end{array} \right] 
\end{equation}

\vspace{2em}
and
\begin{equation}
\mathbf{\hat{M}} = \left[ \begin{array}{*{8}{c@{\hspace{0.18em}}}}
\textbf{m}[1] & \textbf{m}[1] & \textbf{m}[1] & \cdots & \textbf{m}[1] & \cdots  & 0 & 0 \\
0 & \textbf{m}[2]& \textbf{m}[2] & \cdots & \textbf{m}[2] & \cdots  & 0 & 0 \\
0 & 0 & \textbf{m}[3] & \cdots &  \textbf{m}[3]  & \cdots  & 0 & 0 \\
\vdots & \vdots & \vdots & \ddots & \vdots  & \vdots & \vdots & \vdots \\
0 & 0 & 0 & \cdots & 0 & \cdots  &\textbf{m}[{T-2}] & \textbf{m}[{T-2}] \\
0 & 0 & 0 & \cdots & 0 & \cdots  &\textbf{m}[{T-1}] & \textbf{m}[{T-1}] \\
0 & 0 & 0 & \cdots & 0 & \cdots &  0 & \textbf{m}[T]
\end{array} \right] 
\end{equation}

\subsection{Bidirectional Delayed Memory Unit} 
Some non-causal sequence tasks require accessing both past and future input features at a given time. To achieve this, we can utilize a bidirectional PDMU network. This approach efficiently leverages past features (via forward states) and future features (via
backward states) for a specific time frame. The key equation can be expressed as: 
\begin{equation}
\label{pdmu_2}
\mathbf{h}[k]
= \mathbf{m}[k]
+ \sum_{i=k-n}^{k-1} \overset{\rightarrow}{{d}}_{k-i}[i] \, \mathbf{m}[i]
+ \sum_{i=k+1}^{k+n} \overset{\leftarrow}{{d}}_{i-k}[i] \, \mathbf{m}[i]
\end{equation}

In this equation, the delay gates $\overset{\rightarrow}{{d}}_{{k-i}}[i]$ and $\overset{\leftarrow}{{d}}_{{i-k}}[i]$ capture dependencies from past and future memory states, respectively. A partial display of the corresponding $\hat{\textbf{D}}$ matrix is illustrated in Figure \ref{fig:biequation}.
\begin{figure*}[b]
\centering
\resizebox{0.9\textwidth}{!}{
\(
\mathbf{\hat{D}} = \left[ \begin{array}{*{6}{c@{\hspace{0.2em}}}}
1 & \overset{\rightarrow}d_{1}[m] & \overset{\rightarrow}d_2[m] & \cdots & \overset{\rightarrow}d_{n-1}[m] &\overset{\rightarrow} d_{n}[m] \\
\overset{\leftarrow}{d}_1[{m+1}] & 1 & \overset{\rightarrow}d_1[m+1] & \cdots & \overset{\rightarrow}d_{n-2}[m+1] &\overset{\rightarrow} d_{n-1}[m+1] \\
\overset{\leftarrow}{d}_2[m+2] & \overset{\leftarrow}{d}_1[m+2] & 1 & \cdots & \overset{\rightarrow}d_{n-3}[m+2] & \overset{\rightarrow}d_{n-2}[m+2] \\
\vdots & \vdots & \vdots & \ddots & \vdots & \vdots \\
\overset{\leftarrow}{d}_{n-1}[m+n-1] & \overset{\leftarrow}{d}_{n-2}[m+n-1] & \overset{\leftarrow}{d}_{n-3}[m+n-1] & \cdots & 1 & \overset{\rightarrow}d_{1}[m+n-1] \\
\overset{\leftarrow}{d}_{n}[m+n] & \overset{\leftarrow}{d}_{n-1}[m+n] & \overset{\leftarrow}{d}_{n-2}[m+n] & \cdots & \overset{\leftarrow}{d}_{1}[m+n] & 1 \\
\end{array} \right]
\)
}
\caption{Partial display of the bidirectional delay gate matrix $\hat{D}$. }
\label{fig:biequation}
\end{figure*}

\subsection{Efficient Delayed Memory Unit} 
The Efficient PDMU
constrains the PDMU to allow only one delay gate to be active at a time across all channels. This constraint is implemented through a mask operation, succinctly expressed as:

\begin{equation}
    \mathrm {\textbf{Mask}}_i[k] = \begin{cases}
    1, & \text{if } {d}_{i}[k] = \argmax (\textbf{d}[k]) \\
    0, & \text{otherwise}
    \end{cases}
\end{equation}
Then, the output hidden state can be described as:
\begin{equation}
\label{pdmu_3}
\mathbf{h}[k] = \mathbf{m}[k]
+ \sum_{i=k-n}^{k-1} \overset{\rightarrow}{{d}}_{k-i}[i]\mathrm{\mathbf{Mask}}_{k-i}[i]\mathbf{m}[i]
\end{equation}

Due to the non-differentiable nature of the delay gradient, we employ the 'Straight-Through Estimator' (STE) technique \cite{hinton2012neural, bengio2013estimating}.This method addresses the challenge by facilitating optimization via gradient descent. The STE approach allows the forward pass to activate only a single gate while facilitating optimization via gradient descent. Thus, we apply the STE to the output of each layer as follows:

\[
\frac{\partial \mathrm{\textbf{Mask}}_{i}[k]}{\partial {d}_{i}[k]} \approx 1
\]

The training process for the PDMU and its variants, Bi-PDMU and EPDMU, is summarized in Algorithm \ref{alg1}.
\begin{algorithm}[ht]
    \renewcommand{\algorithmicrequire}{\textbf{Input:}}
    \renewcommand{\algorithmicensure}{\textbf{Output:}}
    \caption{Training Procedure for PDMU and Its Variants}
    \label{alg1}
    \begin{algorithmic}[1] 
    \Require Audio/EEG signals at time step \(k\), \(\mathbf{x}[k]\); class label for each task; model hyperparameters.
    \Ensure Optimized Model parameters.
    \State Initialize the Legendre memory unit state-space matrices \(\textbf{A}\) and \(\textbf{B}\), and the delayed memory unit state-space matrices \(\textbf{P}\) and \(\textbf{Q}\) using the Padé approximation according to Equation \ref{pade}.
    \State Discretize the state-space matrices using the zero-order hold (ZOH) method.
    \State Initialize weights and biases using Kaiming uniform initialization.

   \While{the convergence criterion is not met}
    \State Input the signal into the RNN cell as described by Equation \ref{input}.
    \State Compute the memory cell \(\textbf{m}[t]\) as described in Equation \ref{memeoryclee}.
    \State Compute the delay cell and delay gate using Equations \ref{cell_delay} and \ref{delay_cell}.
    \State Compute the output hidden state:
        \If{using PDMU}
            \State Use Equation \ref{pdmu_1}.
        \ElsIf{using Bi-PDMU}
            \State Use Equation \ref{pdmu_2}.
        \ElsIf{using EPDMU}
            \State Use Equation \ref{pdmu_3}.
        \EndIf
    \State Compute the output signal \(\textbf{o}[k]\) using Equation \ref{pmdu_5}.
   \EndWhile
    \end{algorithmic}
\end{algorithm}

\subsection{Spiking Delayed Memory Unit}
Spiking neural networks (SNNs) are energy-efficient owing to their discrete spike activations: a spike realises an accumulation operation, whereas silence incurs no communication. We employ the leaky integrate-and-fire (LIF) neuron \cite{abbott2005model}. The discrete form of the spiking DMU is
\begin{equation} \label{pdmu_spiking} \mathbf{x}[k] = \Theta(\textbf{h}[k]- \theta_{u}) \end{equation}
Here, \(\mathbf{x}[k]\) is the spike vector and \(\Theta(\cdot)\) denotes the Heaviside step function. Whenever the membrane potential exceeds the threshold \(\theta_{u}\), a spike is emitted and \(\mathbf{h}[k]\) is reset to zero. For each branch (memory state and delay-gate state), the input spikes are mapped by the weight matrices \(W_{u}\) and \(W_{v}\), respectively, and then passed through \(\Theta\) to obtain scalar outputs. Thus, \(u[k]\) and \(v[k]\) are binary, taking values in \(\{0,1\}\).

\section{Experimental Results}
\label{experiment}
In our experiment, we endeavor to investigate three primary research questions (RQs):

\textbf{RQ1 (Results):} How does the PDMU model compare in terms of accuracy and effectiveness with other competitive models when processing biosignals that involve both long-term and short-term temporal dependencies?

\textbf{RQ2  (Ablation study):} To what extent does each component of the PDMU contribute to the overall improvements in performance and efficiency?

\textbf{RQ3 (Training time and parameter increments):} Does our model enhance training efficiency and achieve competitive results while requiring fewer parameter increments compared to other models?

To answer the above questions, we evaluate the performance of the proposed model on real-world temporal classification tasks, focusing primarily on cough audio classification, bio-signals (EEG and spiking audio datasets), and two benchmark tasks: speech commands and permuted sequential MNIST. We demonstrate the superiority of the proposed model by comparing it against other state-of-the-art baseline models with similar or fewer neurons.

\subsection{Experimental Setups and Training Configuration}
We selected benchmark tasks in bio-signal processing, spiking audio processing, and speech audio processing to evaluate our model. We implemented all models using the PyTorch library. Across all tasks, we set the nonlinear activation functions $f_u$ and $f_o$ to be $\text{ReLU}(\cdot)$. We used the $\text{softmax}(\cdot)$ function for $\sigma$ to ensure a proper probability distribution. For all experiments, the delay line resolution and discretization resolution, denoted by $\Delta t$, was set to 1 time step.

All models were trained using the Adam optimizer, with a batch size of 128 and a constant learning rate of 0.001. The Kaiming weight initialization approach was adopted for both network weights and biases. The state space matrices $\textbf{A}, \textbf{B}, \textbf{P}, \textbf{Q}$ were discretized using the Padé approximation and the ZOH method. The length of the delay line ranged from 2 to 10, depending on the task. All model training was conducted on Nvidia GeForce GTX 2080Ti GPUs, each equipped with 12 GB of memory.

\subsection{Main results}
\textbf{To address RQ1}, we compared PDMU with several state-of-the-art algorithms across a range of biosignals in healthcare, as well as on benchmarks designed to measure temporal dependencies, such as PS-MNIST.

\subsubsection{Event-based Spoken Word Recognition}
\label{shd}

\begin{figure*}[t]
    \centering
    
    \begin{subfigure}{0.24\linewidth}
        \centering
        \includegraphics[width=\linewidth]{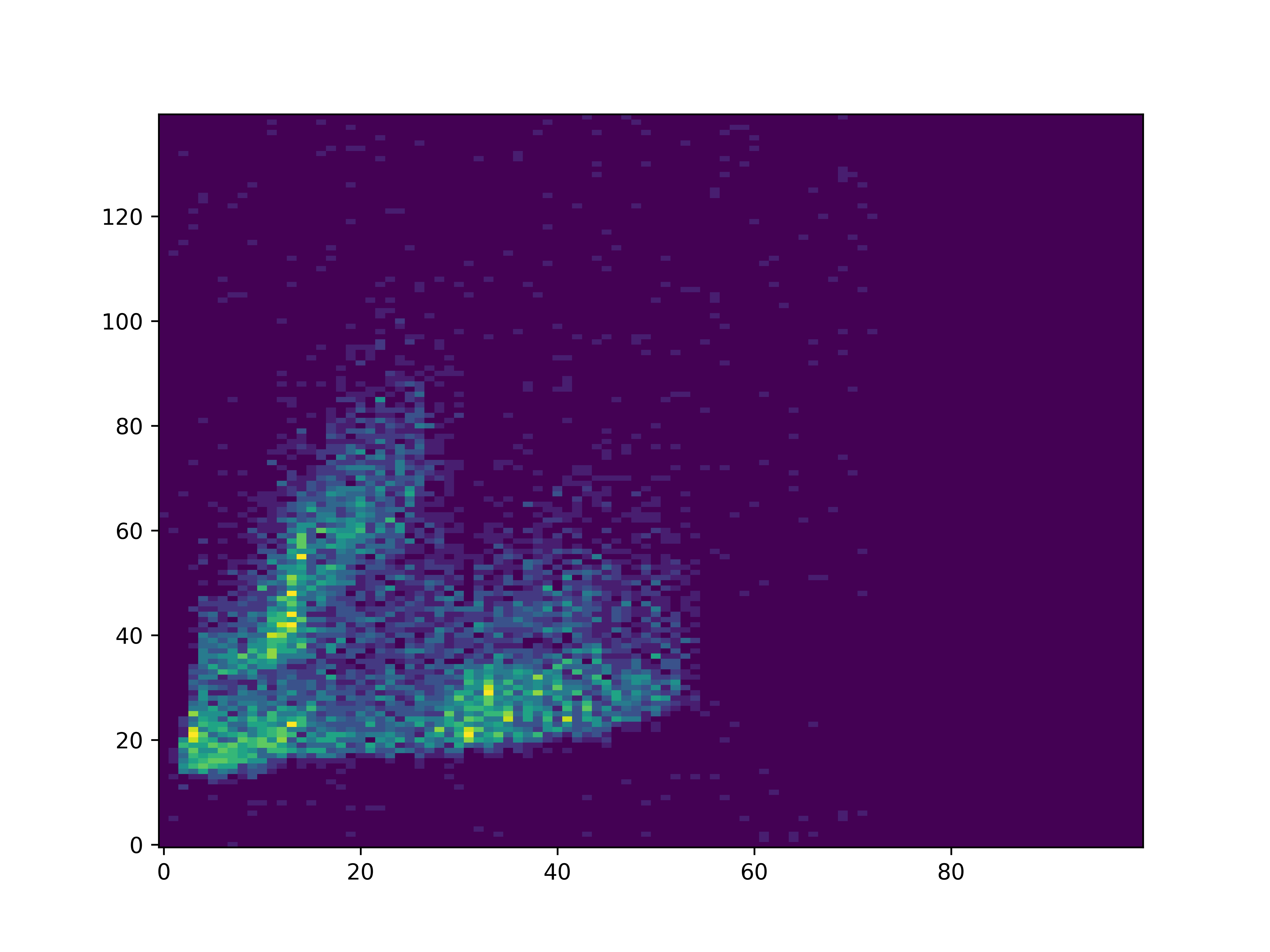}
        \label{fig:subfig1}
    \end{subfigure}%
    \hfill
    \begin{subfigure}{0.24\linewidth}
        \centering
        \includegraphics[width=\linewidth]{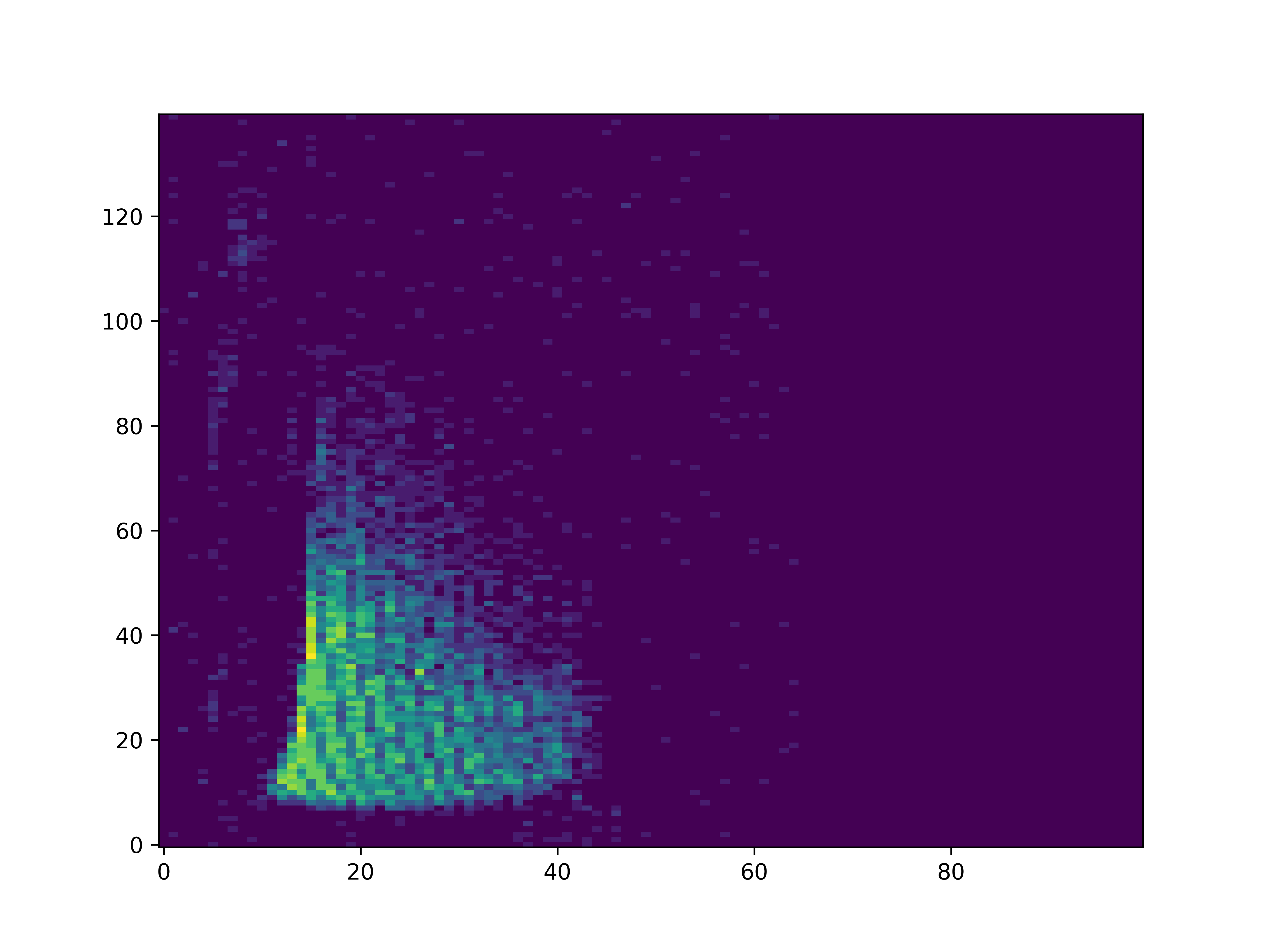}
        \label{fig:subfig2}
    \end{subfigure}%
    \hfill
    \begin{subfigure}{0.24\linewidth}
        \centering
        \includegraphics[width=\linewidth]{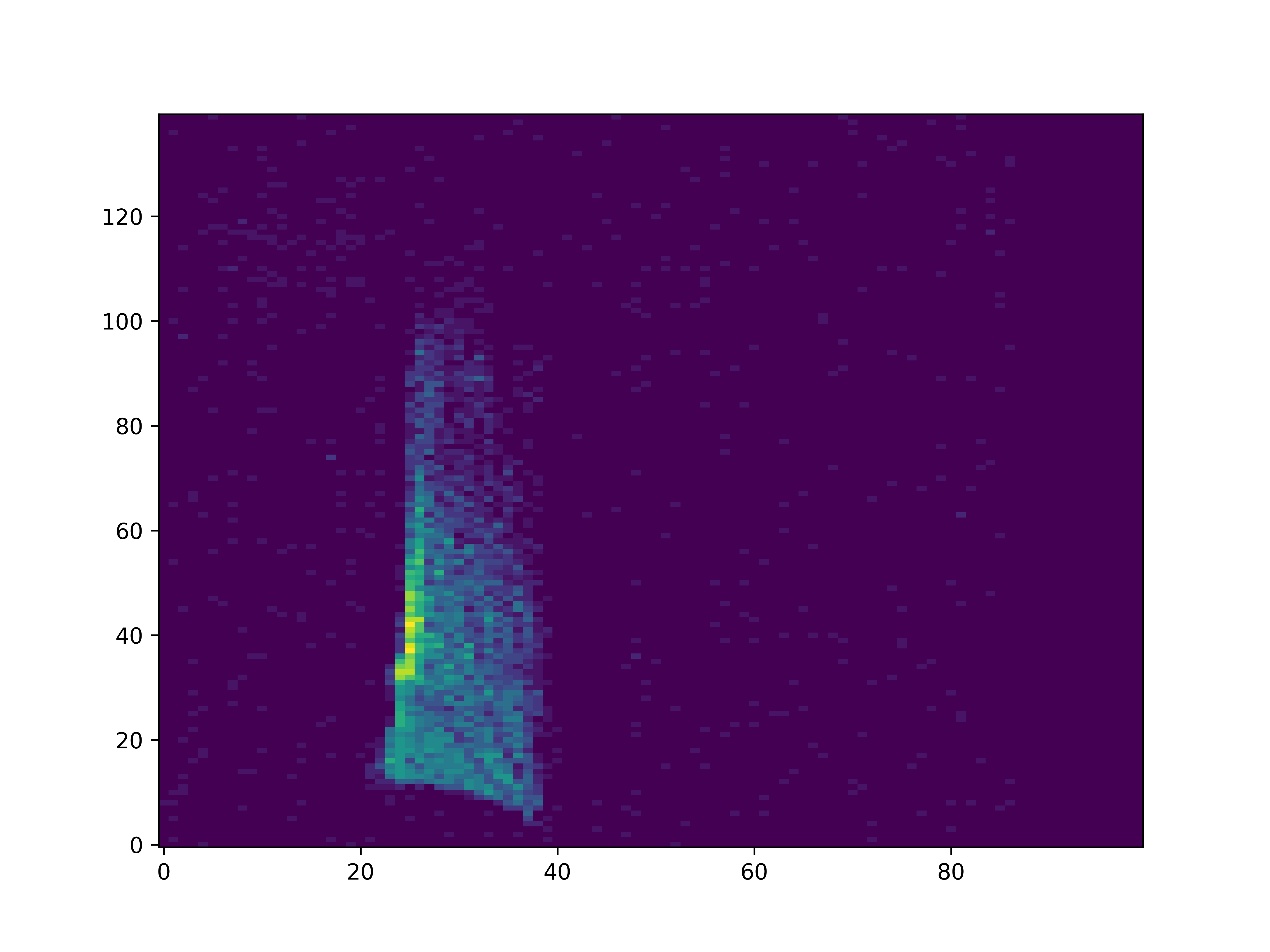}
        \label{fig:subfig3}
    \end{subfigure}%
    \hfill
    \begin{subfigure}{0.24\linewidth}
        \centering
        \includegraphics[width=\linewidth]{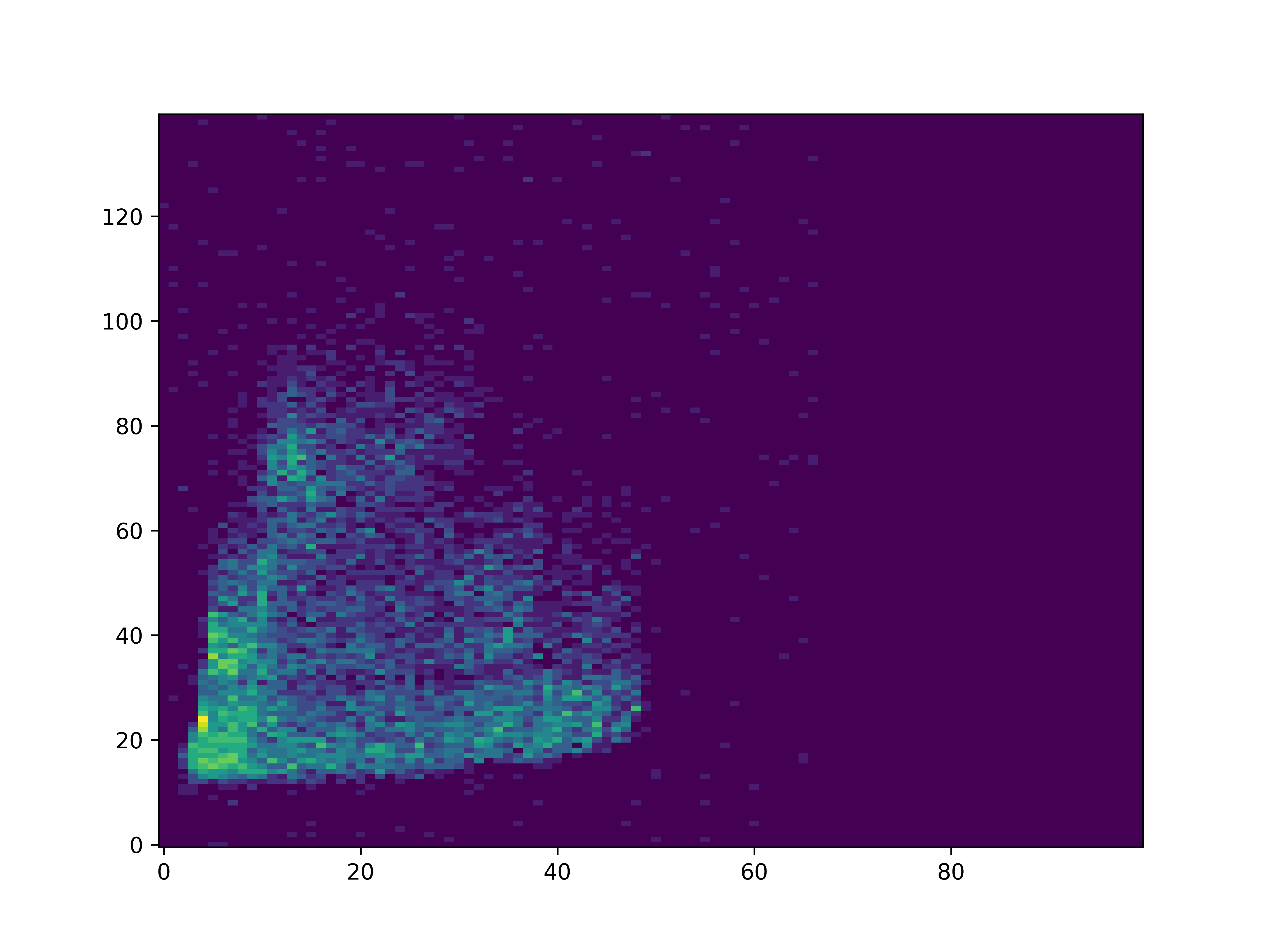}
        \label{fig:subfig4}
    \end{subfigure}

    \vspace{2mm} 

    \begin{subfigure}{0.24\linewidth}
        \centering
        \includegraphics[width=\linewidth]{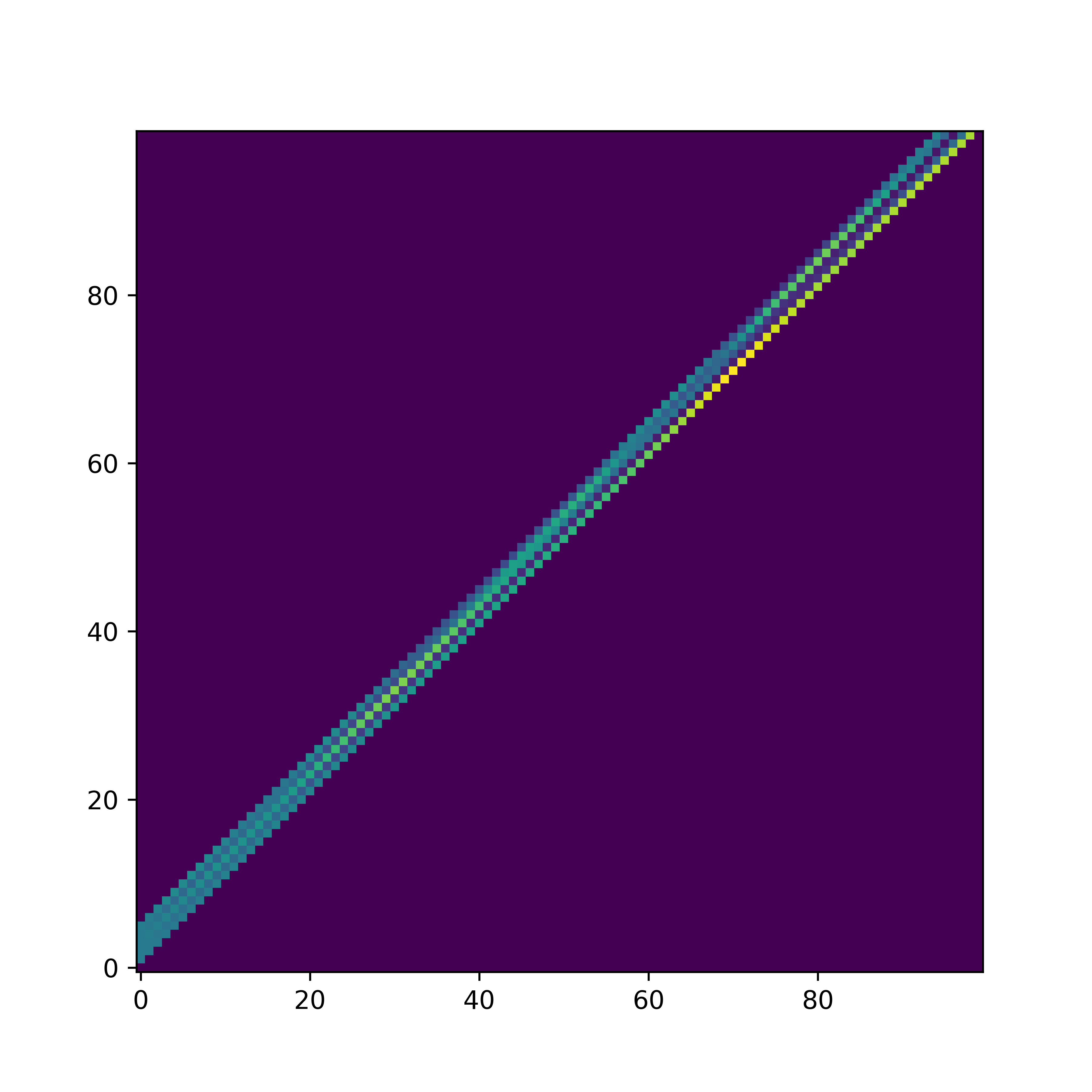}
        \caption{Voice '1'}
        \label{fig:subfig5}
    \end{subfigure}%
    \hfill
    \begin{subfigure}{0.24\linewidth}
        \centering
        \includegraphics[width=\linewidth]{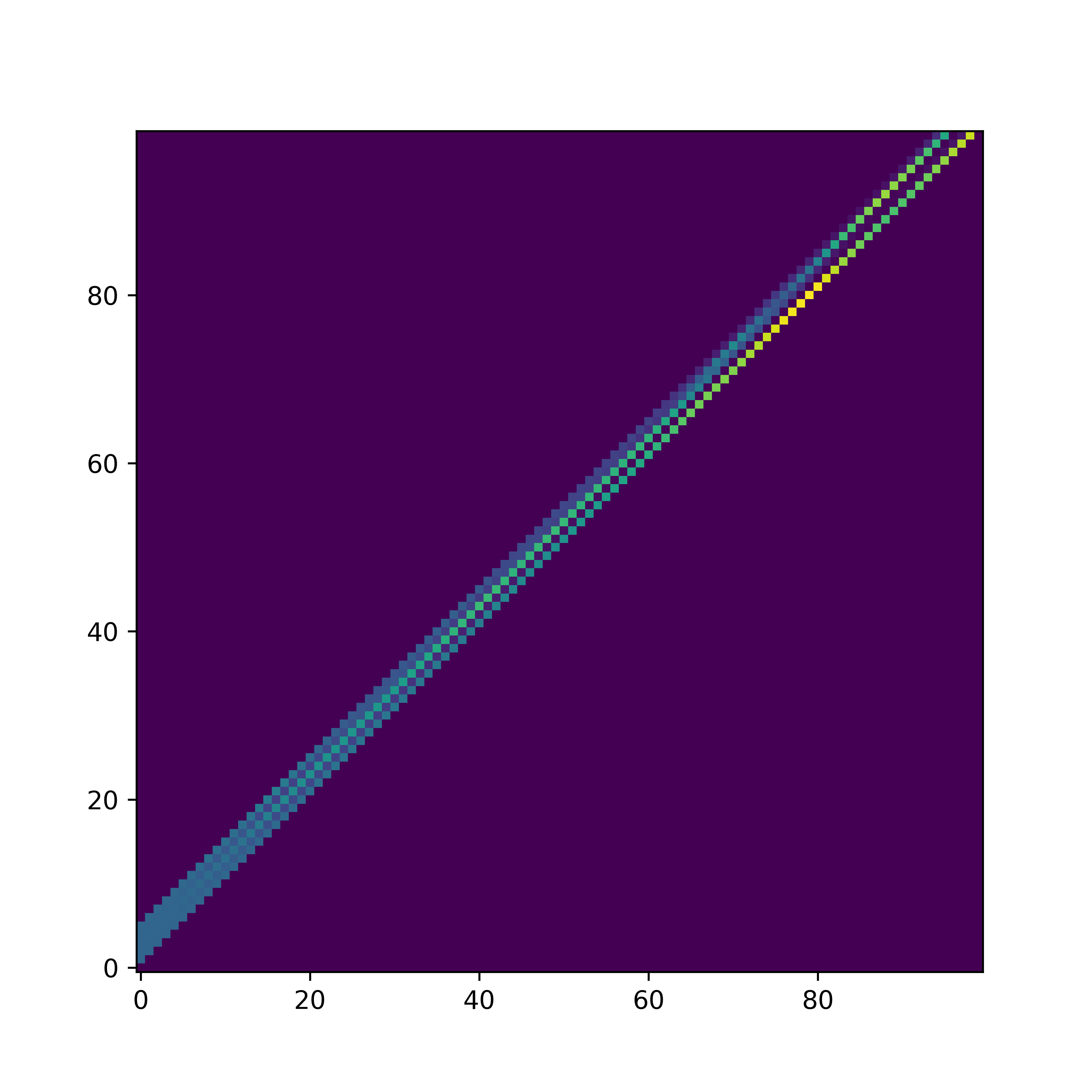}
        \caption{Voice '2'}
        \label{fig:subfig6}
    \end{subfigure}%
    \hfill
    \begin{subfigure}{0.24\linewidth}
        \centering
        \includegraphics[width=\linewidth]{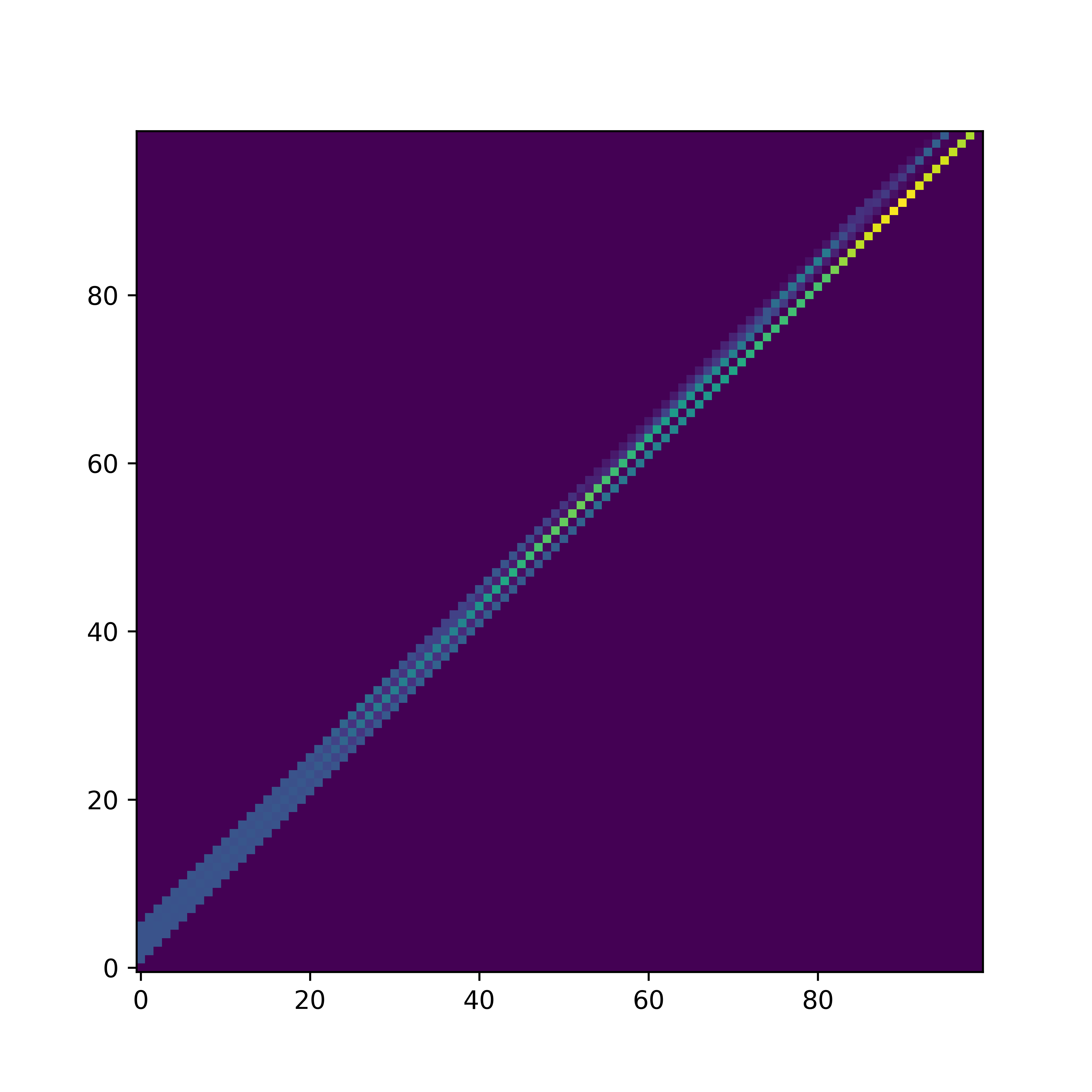}
        \caption{Voice '6'}
        \label{fig:subfig7}
    \end{subfigure}%
    \hfill
    \begin{subfigure}{0.24\linewidth}
        \centering
        \includegraphics[width=\linewidth]{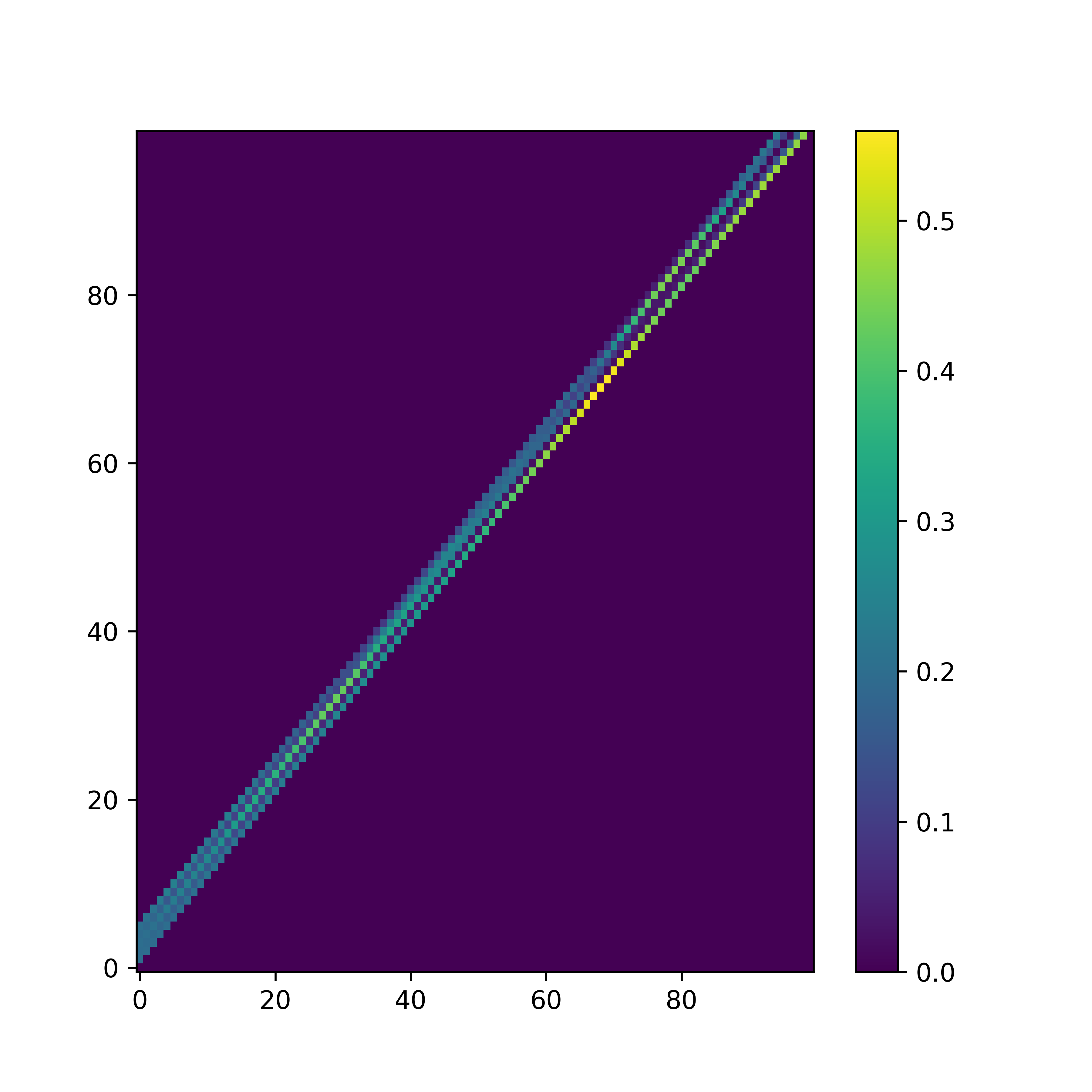}
        \caption{Voice '9'}
        \label{fig:subfig8}
    \end{subfigure}

    \caption{Visualization of SHD  Samples and Hidden Layer Attention Maps.
Panels (a) voice '1', (b) voice '2', (c) voice '6', and (d) voice '9' are shown. The top row illustrates the input samples (x-axis: time steps; y-axis: channels), while the bottom row shows the corresponding gating maps at the hidden layer (both axes denote time steps).}
    \label{fig:attentionall}
\end{figure*}

The Spiking Heidelberg Dataset (SHD) \cite{cramer2020heidelberg} was introduced for event-based speech processing research. This dataset is generated by converting raw audio waveforms into an event-based representation using a biologically plausible artificial cochlear model. In this study, we utilize the SHD dataset to evaluate the applicability and generalization of the PDMU across various data modalities. Each SHD sample consists of 700 channels, each representing distinct frequency-sensitive neurons in the peripheral auditory system. To ensure adequate temporal resolution and minimize post-processing effort, spikes are aggregated within 10 ms time bins, resulting in 100 time steps. 

{We employ a two-layer neural network, using the states (or membrane potential) at the last time step for decoding. We employ this decoding method to assess the network's long-term memory capabilities which follows the work by previous RNNs. In this task, delay-based models have demonstrated state-of-the-art performance. For example, models incorporating temporal attention and learnable axonal delays have achieved cutting-edge results on the SHD dataset \cite{tempoaralatten, sun2023learnable, hammouamrilearning,d2024denram,patino2024hardware,malettira2024tskips, meszaros2025efficient,queant2025delrec}. Furthermore, models using parameter-free attention mechanisms \cite{sun2025towards} have reached accuracies up to 96.26\%. However, these approaches typically rely on Spikerate decoding methods \cite{shrestha2022spikemax} and employ larger network architectures and introduce decision latency, which are not compatible with our current framework. Furthermore, as illustrated in Figure \ref{fig:attentionall}, we present the attention maps generated by the gating mechanism. When the input contains substantial information, as seen in panels (a) to (c), where the spike trains are dense, the gating function exhibits complex and dynamic interactions. In contrast, when the input carries minimal information, the gating mechanism functions analogously to a skip connection, assigning the highest attention weights to the preceding states \cite{cook2025brain}. This allows information from earlier time steps to be transmitted forward with minimal attenuation, effectively preserving historical signals. }

\begin{table}[h]
\centering
\normalsize
\caption{Comparison of model performance on the SHD dataset. The architecture utilized two layers, each with 128 neurons. The Skip DMU$^*$ denotes a single constant-delay gate (skip connection in the temporal domain). }
\label{tbl:shd}
\begin{tabular}{{lcc}}
    \cline{1-3}
    \multicolumn{1}{l}{\textbf{Methods}} &
    \multicolumn{1}{c}{\textbf{SNN}} &
    \multicolumn{1}{c}{\textbf{Accuracy}} \\
    \hline
    FSNN \cite{cramer2020heidelberg}   &Yes& 48.10\% \\
    RSNN \cite{cramer2020heidelberg} &Yes& 83.20\% \\
    Adaptive RSNN \cite{yin2020effective}  &Yes& 84.40\% \\
     TC-LIF (Rec)\cite{zhang2024tc}   &Yes& 88.91\% \\
    \textbf{Spiking DMU}  &Yes& \textbf{90.98\%} \\
    \hline
    LSTM\cite{yin2021accurate}  &No & 79.90\% \\
      Skip DMU$^*$   &Yes& 82.41\% \\ 
    \textbf{PDMU} & No& 84.57\%  \\
    \textbf{EPDMU} & No & \textbf{86.48}\%  \\
    \textbf{Bi-PDMU} & No & 84.94\%  \\
    \hline
\end{tabular}
\end{table}

The results of our model, along with other state-of-the-art recurrent spiking neural networks (RSNNs) and LSTM models, are provided in Table \ref{tbl:shd}. Notably, for the spiking models, our proposed Spiking DMU model surpasses RSNN models by 7.78\%. This result also competes favorably with the state-of-the-art RSNN model that employs temporal attention and advanced data augmentation techniques \cite{yao2021temporal}, as well as the two-component neuron models \cite{zhang2024tc}. {In addition, we evaluated an architecture equipped with a skip connection, which achieved an accuracy of 82.41\%. Although this performance is an improvement over the baseline, it remains lower than that of the delay-line architecture, underscoring the importance of the internal complex interactions. }

\begin{table}[H]
\normalsize
\centering
\caption{Comparison of model performance on COUGHVID dataset.  }
\label{tbl:coughvid}
\begin{tabular}{lcccc}
    \hline
    \textbf{Methods } & \textbf{SNN} & \textbf{Accuracy}& 
    \textbf{AUC} &
        \textbf{F1} \\
    \hline
    LSTM \cite{hamdi2022attention} & No & 76.41\% & 76.26\%&   74.48\%  \\
    
    \textbf{PDMU} & No & 82.45\% & 81.58\%& 82.21\%\\
    \textbf{EPDMU} & No & \textbf{82.45\%}& 82.08\%& 82.32\%\\
    \textbf{Bi-PDMU} & No & 79.63\% &78.02\%& 78.23\%\\
        \textbf{Spiking-DMU} & Yes & 82.37\% &81.88\%& 82.01\%\\
        \hline
CNN  & No & 81.37\% &82.27\% &80.97\%\\
    \textbf{CNN-PDMU} & No & \textbf{85.48}\% &84.13\%& 85.10\%\\       
    \hline
\end{tabular}

\end{table}

\subsubsection{Cough Audio Signal Classification}
\label{cough}
The COUGHVID dataset, developed by Orlandic et al.\cite{orlandic2021coughvid}, is a large-scale, publicly available corpus consisting of 27,550 cough recordings. Collected between April 1 and December 1, 2020, via a web application, participants provided metadata on age, gender, geolocation, pre-existing respiratory conditions, and COVID-19 status (Healthy, COVID-19, Symptomatic). The recordings, encoded in Opus at a 48kHz sampling rate, were annotated by four physician experts for cough quality, type, and other audible symptoms. The raw audio data was transformed into Mel spectrograms, each with 128 channels, and analyzed over simulation time steps of 420 (approximately 9.75 seconds).

A comparative analysis of models on the COUGHVID dataset (see Table \ref{tbl:coughvid}) demonstrates that CNN-PDMU achieved the accuracy of 85.48\%. The pure CNN model (replacing the PDMU module with fully connected layers) only attained an accuracy of 81.37\%, indicating the importance of capturing temporal dependencies in this dataset. Both PDMU and EPDMU performed well, each with an accuracy of 82.45\%. The LSTM model achieved an accuracy of 76.41\%, while Spiking-DMU reached 82.37\%, highlighting the potential of neuromorphic computing in cough classification. {In addition, following the similar research \cite{conguh}, we also report the F1 and AUC metrics. The improvements persist under F1/AUC, indicating that the gains stem from better discriminative capacity rather than class imbalance. } This study highlights the effectiveness of our model in analyzing cough sounds for diagnostic purposes.

\subsubsection{EEG Attention Detection}
\label{eeg}
The WithMe dataset, derived from the WithMe experiment \cite{de2022no, sun2024eeg,sun2025electroencephalography}, includes EEG recordings from 42 participants. For this study, we selected 38 participants for training and internal testing, dividing their data into training and testing sets, referred to as within-subjects. Data from the remaining 4 participants were used to evaluate our model's generalizability, termed unseen-subjects. Specifically, we partitioned the WithMe data into a training set and two testing sets: one for within-subject evaluations, consisting of 18,176 training instances and 4,580 validation instances, and another for unseen-subject assessments, with 2,400 validation instances. Preprocessing the EEG data involved re-referencing each channel to the average activity of the mastoid electrodes, band-pass filtering between 1 and 30 Hz, and downsampling to 64 Hz. The data were then segmented into 1.2-second epochs based on trigger events. The final preprocessing step normalized the EEG channel data to ensure zero mean and unit variance for each sample. The dataset is available for access at the following link 
\footnote{https://github.com/sunpengfei1122/Withme-EEG-dataset}.

\begin{table}[]
\centering
\normalsize
\caption{Comparison of model performance on the WithMe dataset. The Accuracy column shows performance results, with the former indicating within-subject performance and the latter showing unseen-subject performance. The architecture utilized two layers, each with 64
neurons.}
\label{tbl:withme}
\begin{tabular}{{lcc}}
    \cline{1-3}
    \multicolumn{1}{l}{\textbf{Methods}} &
        \multicolumn{1}{c}{\textbf{SNN}} &
    \multicolumn{1}{c}{\textbf{Accuracy}} \\
    \hline
    EEGNet \cite{lawhern2018eegnet}  & No& 81.67\%/76.42\% \\
    LSTM & No & 80.09\%/74.00\% \\
    DMU \cite{sun2024eeg} & No& 81.94\%/75.92\% \\
    \textbf{PDMU} & No & 84.17\%/74.50\% \\
    \textbf{EPDMU}  & No &84.43\%/74.58\% \\
    \textbf{Bi-PDMU}  &No& \textbf{84.82\%/77.13\%} \\
    \cline{1-3}
    \textbf{Spiking DMU}  &Yes& 80.21\%/75.08\% \\ 
    \hline

\end{tabular}
\end{table}

As shown in Table \ref{tbl:withme}, our proposed model outperforms other baseline models. Compared to LSTM, it surpasses within-subject performance by 4.08\% and unseen-subject performance by 0.5\%. Furthermore, with the introduction of the bidirectional delay gate, it achieves the best accuracy of 77.13\% on unseen subjects. For the spiking DMU, it achieves slightly better performance compared to the LSTM while requiring much fewer parameters and offering energy efficiency due to its sparse activations. These results underscore the effectiveness of the proposed method and its superior temporal modeling capability.

\subsubsection{Speech Processing}
\label{sc}
The Speech Commands V2 dataset \cite{warden2018speech} consists of 105,829 audio files featuring 35 different words. Each recording lasts up to 1 second and is sampled at 16 kHz. The dataset is divided into 84,843 samples for training, 9,981 for validation, and 11,005 for testing. For our study, we selected the most challenging classification task involving all 35 classes to comprehensively assess our models' performance.

As reported in Table \ref{tbl:gsc}, our spiking DMU achieves the best results among spiking neural networks, with a noteworthy performance of 96.39\%. For the non-spiking version, we compare our method with other models that can perform sequential inference, demonstrating competitive results. {There is a $\sim$1\% gap between our model and the transformer-based SOTA~\cite{sc1}. However, our model uses far fewer parameters (1.6M vs.\ 86.9M; $\approx 53\times$ fewer) and operates causally with strictly sequential inference. Relative to UniRepLKNet~\cite{UniRepLKNe}, the accuracy gap is $\sim$1.5\% while the parameter count remains much smaller (1.6M vs.\ 55.5M; $\approx 34\times$ fewer).}

\begin{table}[t]

\centering
\normalsize
\caption{Comparison of model performance on GSC dataset.}
\begin{tabular}{lcc}

\label{tbl:gsc}
    \textbf{Methods } & \textbf{SNN} & \textbf{Accuracy} \\
    \hline
    Rate-based SNN \cite{yilmaz2020deep} & Yes & 75.20\% \\
    SRNN+ALIF \cite{yin2021accurate}  & Yes & 92.10\% \\
    SNN \cite{salaj2021spike}  & Yes & 89.04\% \\
    SNN with SFA \cite{salaj2021spike} & Yes & 91.21\% \\
    LIF \cite{bittar2022surrogate}  & Yes & 83.03\% \\
    RadLIF \cite{bittar2022surrogate} & Yes & 94.51\% \\
    TC-LIF \cite{zhang2024tc} & Yes & 94.84\% \\
    \textbf{Spiking DMU} & Yes & \textbf{96.39}\% \\
    \hline
    RNN \cite{bittar2022surrogate} & No & 92.09\% \\
    liBRU \cite{de2018neural}& No & 95.06\% \\
    LMUFormer \cite{liulmuformer}& No & 96.53\% \\
    \textbf{PDMU} & No & 96.93\% \\
    \textbf{EPDMU} & No & \textbf{97.01\%} \\
    \textbf{Bi-PDMU} & No & 96.96\% \\
    \hline
\end{tabular}
\end{table}

\begin{table}[t]
\vspace{0.5cm}
\centering
\small
\caption{Comparison of model performance on PSMNIST dataset.}
\label{tbl:psmnist}
\begin{tabular}{@{}l@{}c@{}c@{}c@{}}
    \cline{1-4}
    \multicolumn{1}{l}{\textbf{Methods}} &
    \multicolumn{1}{c}{\textbf{Architecture}} &
    \multicolumn{1}{c}{\textbf{SNN}} &
    \multicolumn{1}{c}{\textbf{Accuracy}} \\
    \hline
    GRU \cite{chandar2019towards} & - &No & 92.39\% \\
    LSTM \cite{chandar2019towards} & 200 &No & 89.86\% \\
    Lipschitz RNN \cite{erichson2020lipschitz} & 200 &No & 96.40\% \\
    coRNN \cite{rusch2020coupled} & 200 &No & 96.06\% \\
    DMU \cite{sun2023delayed} & 200 &No & 96.39\% \\
       LMU\cite{voelker2019legendre} & 200+200 &No & 97.15\%  \\
    \textbf{PDMU} & 200+200 &No & 98.22\%  \\
    \textbf{EPDMU} & 200+200 &No & 98.24\%  \\
    \textbf{Bi-PDMU} & 200+200&No  & \textbf{98.25}\%  \\
    \hline
   SRNN+ALIF \cite{yin2021accurate} & 64+256+256 &Yes & 94.30\%  \\
    
    TC-LIF \cite{zhang2024tc} & 64+256+256&Yes & 95.36\%  \\
    \textbf{Spiking DMU} & 200+200&Yes  & \textbf{96.07}\%  \\
    \hline
\end{tabular}
\end{table}

\subsubsection{Permuted Sequential Image Classification}
\label{psmnist}
The permuted sequential MNIST (PSMNIST) dataset \cite{le2015simple} was developed to assess the capability of RNN models in capturing long-range temporal dependencies. This dataset is derived from the MNIST dataset by first flattening each image into a 784-dimensional vector and then shuffling its spatial information using a consistent permutation vector. In our experiments, the elements of this vector are presented sequentially to the network, with decisions made after processing all pixels. This task requires the model to reconstruct the original temporal order and identify dependencies between different pixels.

Results in Table \ref{tbl:psmnist} show that our proposed PDMU competes with other RNN models engineered for learning long-term dependencies, such as Lipschitz RNN and coRNN \cite{rusch2021long, rusch2020coupled, erichson2020lipschitz}. {Our model attains 98.22\% with around 0.2M parameters, compared with 98.84\% reported by 2M parameter CNN designed for long-range dependencies~\cite{psmnist}. The 0.62\% accuracy gap is modest relative to the  10 $\times$ reduction in model size.
}. Furthermore, the spiking DMU achieves a test accuracy of 96.07\% using only two spiking layers with 200 neurons,  establishing a new benchmark for this task.

\subsection{Ablation study}

\begin{table*}[t]
\centering
\normalsize
\caption{Ablation study on different tasks including EEG classification (WithMe), audio processing (GSC and SHD), and PSMNIST. The GSC$^*$ represents the LMU cell augmented with LMUFormer\cite{liulmuformer}. }
\label{tbl:ablation}
\begin{tabular}{lccccc}
    \hline
    \multicolumn{1}{l}{\textbf{Datasets}} &
    \multicolumn{1}{c}{\textbf{DMU}} &
    \multicolumn{1}{c}{\textbf{SNN}} &
        \multicolumn{1}{c}{\textbf{Parallel
Training}} &
    \multicolumn{1}{c}{\textbf{Acc. (\%)}} &
    \multicolumn{1}{c}{\textbf{$\Delta$ (\%)}} \\
    \hline
    \multirow{4}{*}{COUGHVID} &No & No &Yes& 80.23 & 0.00\\
    &Yes & No &Yes& 82.45 & \textbf{$\uparrow${2.22} }\\
    \cline{2-6}
     &No & Yes &No& 76.52 & 0.00  \\
    &Yes & Yes &No&  82.37& \textbf{$\uparrow$5.85}\\
    \hline
    
    \multirow{4}{*}{WithMe} &No & No &Yes& 82.68/69.83 & 0.00/0.00\\
    &Yes & No &Yes& 84.17/74.50 & \textbf{$\uparrow$1.49}/\textbf{$\uparrow$4.67}\\
    \cline{2-6}
     &No & Yes &No& 79.24/66.83 & 0.00/0.00 \\
    &Yes & Yes &No& 80.21/75.08 & \textbf{$\uparrow$0.97}/\textbf{$\uparrow$8.25}\\
    \hline
    \multirow{4}{*}{SHD} &No & No&Yes & 78.16 & 0.00 \\
    &Yes & No &Yes& 84.57 & \textbf{$\uparrow$6.41}\\
    \cline{2-6}
     &No & Yes&No & 85.09 & 0.00 \\
    &Yes & Yes &No& 90.98 & \textbf{$\uparrow$5.89}\\
    \hline
        \multirow{4}{*}{PSMNIST} &No & No&Yes & 97.85 & 0.00 \\
    &Yes & No&Yes & 98.22 &\textbf{$\uparrow$0.37} \\
    \cline{2-6}
     &No & Yes &No& 93.75 & 0.00 \\
    &Yes & Yes &No& 96.07 &\textbf{ $\uparrow$2.32}\\
    \hline
        \multirow{3}{*}{GSC} &No & No&Yes & 74.44 & 0.00 \\
    &Yes & No &Yes& 78.52 & \textbf{$\uparrow$4.08}\\
    &Yes (Bi) & No &Yes& 80.07 & \textbf{$\uparrow$5.63}\\
    \hline
            \multirow{4}{*}{GSC$^*$} &No & No&Yes & 96.53 & 0.00 \\
    &Yes & No&Yes & 96.93 &\textbf{$\uparrow$0.40} \\
    \cline{2-6}
     &No & Yes&No & 95.71 & 0.00 \\
    &Yes & Yes&No & 96.39 & \textbf{$\uparrow$0.68}\\ 
    \hline
\end{tabular}
\end{table*}

\textbf{To answer RQ2}, {we compare our approach with the baseline LMU  module and calculate the performance improvement of each variant. The results, presented in Table \ref{tbl:ablation}, demonstrate that the PDMU model, when enhanced with our delayed memory unit, achieves significant improvements across all tasks. Specifically, in the non-spiking version, the causal delayed memory unit exhibits substantial gains in tasks requiring short-term memory, such as EEG classification and spiking spoken word detection. This suggests that while the LMU can capture very long-term memory, it may overlook detailed features present in bio-signals.  Gains are largest on tasks with prominent short-term structure: on SHD, +6.41\% (non-spiking) and +5.89\% (spiking); on WithMe, +1.49\% (within-subject) and +4.67\% (unseen-subject) in the non-spiking setting, and +0.97\% / +8.25\% for the spiking setting. On GSC, PDMU yields +4.08\%, while Bi-PDMU reaches +5.63\%, highlighting the benefit of improved temporal credit assignment. When the baseline is already strong (PSMNIST and GSC$^*$), improvements are smaller but consistent (typically +0.37\% to +0.68\%). Overall, these trends support our design goal: the delay-gated path complements the LMU’s long-range memory by selectively aggregating recent states, yielding robust gains with minimal overhead.}

\begin{figure}[h]
    \centering
  \includegraphics[width=1\linewidth]{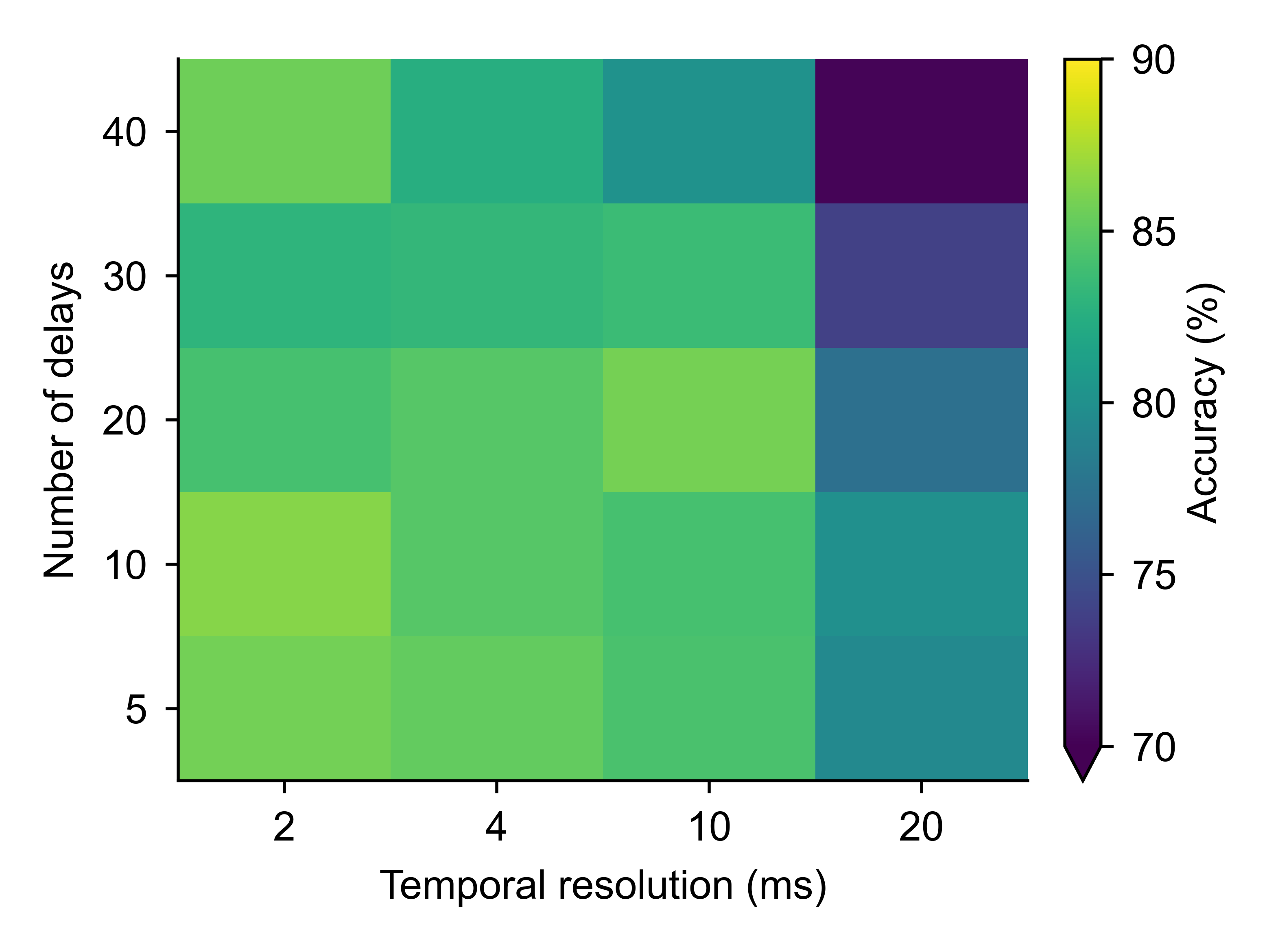}
    \caption{{ Landscape showing classification accuracy as a function of temporal resolution (x-axis) and number of delays
(y-axis) across SHD dataset.}}
    \label{fig:gate1}
    
\end{figure}
\begin{figure*}[!h]
    \centering
\includegraphics[width=0.90\linewidth]{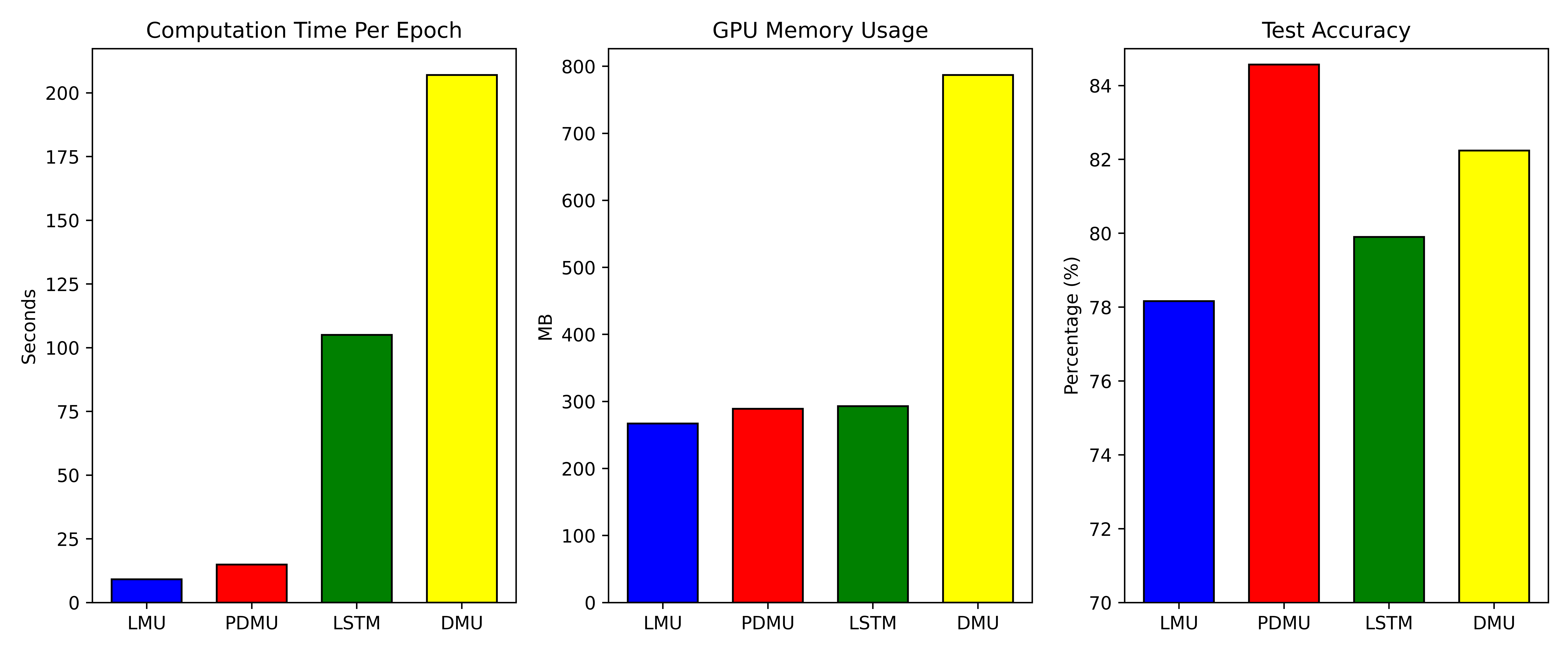}
    \caption{Comparison of the computation time, GPU memory usage, and test accuracy of the proposed PDMU over other RNNs on the SHD dataset. }
    \label{fig:computation}  
\end{figure*}

\subsection{Training time and parameter increments}

\textbf{To answer RQ3}, {we first evaluated both the temporal resolution (time-step size) and the number of delays on SHD. Across settings (Figure~\ref{fig:gate1}), performance is stable, and a moderate delay budget (about 10 delays) consistently works well. Very long delay lines degrade accuracy by injecting stale context (e.g., at resolution $=20$ with 50 simulation steps, using 40 delays reduces accuracy by $9\%$). In practice, the base SSM captures smoother, long-range context, while the delay path emphasizes short transients; excessive delays blur this balance. For this dataset, a total delay of less $100 \mathrm{ms}$ is a sensible target. Since increasing the number of delays also increases model size, we keep the delay count below 10 across all datasets (typically 5).}

We evaluate the learning efficiency and GPU resource utilization of our proposed method compared to other recurrent neural network architectures, including the Legendre Memory Unit (LMU), Long Short-Term Memory (LSTM), and Delayed Memory Unit (DMU). Experiments were performed on a single Nvidia GeForce GTX 2080Ti GPU, utilizing a network configuration comprising two layers, each with 128 neurons, and a batch size of 64.

As illustrated in Figure \ref{fig:computation}, when processing the Spiking Heidelberg Digits (SHD) dataset, both the LSTM and DMU exhibit extended computation times due to their need to store and process intermediate neuronal states. In contrast, our model benefits from parallel training, achieving a 9 $\times$ and 29 $\times$ training speed up compared to the LSTM and DMU, respectively, while also delivering superior test performance. Regarding GPU memory consumption, the DMU shows the highest usage, primarily due to its requirement for a sliding memory mechanism to maintain delayed hidden states.

Regarding parameter efficiency, the PDMU is designed as a streamlined, state–space auxiliary module with learnable delay gates (typically fewer than 10). In the illustrative configuration where the delay line length is set to 5, the gate state–space uses two matrices \(\textbf{P}\in\mathbb{R}^{5\times 5}\) and \(\textbf{Q}\in\mathbb{R}^{5\times 1}\), contributing \(25+5=30\) parameters. In addition, two small linear mappings handle the input signal and the delay gate; their contributions are modest (on the order of the input and delay dimensions).

{ At inference time, all three models are causal RNN variants and thus have the same single-
step latency (real time). However, their per-sample compute differs because the intermediate
activation and gate counts are different. We thus report (i) the total parameter count per layer and (ii) the amount of \emph{runtime state} per layer that hardware must retain between steps as a proxy for compute comparison. Let \(M\) denote the input dimension, \(N\) the number of output neurons, and \(d\) the number of delays (typically \(d<10\)). Then, for \emph{inference parameter counts}:
\begin{align*}
\text{LSTM:}\quad
& \#\text{params} = 4\big(MN + N^{2} + N\big), \\
\text{DMU:}\quad
& \#\text{params} = MN + N^{2} + N \;+\; Md \;+\; d^{2} \;+\; d, \\
\text{PDMU:}\quad
& \#\text{params} = N^{2} + M + N \;+\; 2MN \;+\; d^{2} \;+\; d.
\end{align*}
}
{For \emph{inference resources}, we have:
\begin{align*}
\text{LSTM:}\quad
& \#\text{runtime states} = 2N \quad \\
\text{DMU:} \quad & \#\text{runtime states} = N \times d \\
\text{PDMU:}\quad
& \#\text{runtime states} = N + d
\end{align*}}

{Thus, PDMU adds only a small overhead relative to a standard State\mbox{-}Space Model (SSM), is smaller than LSTM (which carries the four-gate factor), and, unlike DMU, does not require an \(N \times d\) buffer to store delayed hidden states, because PDMU compresses past information into a vector via the state–space module. }

\section{Conclusion}

\label{discussion}
In this study, we introduced the Parallel Delayed Memory Unit (PDMU), an RNN variant designed to enhance the efficiency and effectiveness of audio and biomedical signal processing. The PDMU incorporates a delay line structure with delay gates to improve short-term temporal interaction and credit assignment, supporting parallel training and sequential inference.

Our experiments demonstrate that the PDMU achieves competitive performance on benchmarks such as cough audio signal classification, EEG detection, spiking spoken word detection, speech processing, and permuted sequential image classification. Variants like bi-directional and efficient PDMUs further optimize performance and energy efficiency. The PDMU's ability to process complex biomedical and audio data efficiently has significant implications for medical diagnostics and explorations, facilitating advancements in health and medicine through improved signal analysis. {In addition, the module is plug-and-play: the delay mechanism is, in principle, compatible with any SSM framework, and we plan to explore such integrations in future work.}

\bibliographystyle{unsrt} 
\bibliography{sample}{}

\begin{thebibliography}{10}

\bibitem{ravanelli2017improving}
Mirco Ravanelli, Philemon Brakel, Maurizio Omologo, and Yoshua Bengio.
\newblock Improving speech recognition by revising gated recurrent units.
\newblock {\em arXiv preprint arXiv:1710.00641}, 2017.

\bibitem{sun2023delayed}
Pengfei Sun, Jibin Wu, Malu Zhang, Paul Devos, and Dick Botteldooren.
\newblock Delayed memory unit: Modeling temporal dependency through delay gate.
\newblock {\em IEEE Transactions on Neural Networks and Learning Systems},
  2024.

\bibitem{fernando2022deep}
Tharindu Fernando, Harshala Gammulle, Simon Denman, Sridha Sridharan, and
  Clinton Fookes.
\newblock Deep learning for medical anomaly detection-a survey.
\newblock {\em ACM Comput. Surv.}, 54(7):141--1, 2022.

\bibitem{qian2024learning}
Kun Qian, Zhihao Bao, Zhonghao Zhao, Tomoya Koike, Fengquan Dong, Maximilian
  Schmitt, Qunxi Dong, Jian Shen, Weipeng Jiang, Yajuan Jiang, et~al.
\newblock Learning representations from heart sound: A comparative study on
  shallow and deep models.
\newblock {\em Cyborg and Bionic Systems}, 5:0075, 2024.

\bibitem{vaiciukynas2017detecting}
Evaldas Vaiciukynas, Antanas Verikas, Adas Gelzinis, and Marija Bacauskiene.
\newblock Detecting parkinson’s disease from sustained phonation and speech
  signals.
\newblock {\em PloS one}, 12(10):e0185613, 2017.

\bibitem{de2022no}
Jorg De~Winne, Paul Devos, Marc Leman, and Dick Botteldooren.
\newblock With no attention specifically directed to it, rhythmic sound does
  not automatically facilitate visual task performance.
\newblock {\em Frontiers in Psychology}, 13:894366, 2022.

\bibitem{hochreiter1997long}
Sepp Hochreiter and J{\"u}rgen Schmidhuber.
\newblock Long short-term memory.
\newblock {\em Neural computation}, 9(8):1735--1780, 1997.

\bibitem{vaswani2017attention}
Ashish Vaswani, Noam Shazeer, Niki Parmar, Jakob Uszkoreit, Llion Jones,
  Aidan~N Gomez, {\L}ukasz Kaiser, and Illia Polosukhin.
\newblock Attention is all you need.
\newblock {\em Advances in neural information processing systems}, 30, 2017.

\bibitem{9844844}
Guochen Yu, Andong Li, Hui Wang, Yutian Wang, Yuxuan Ke, and Chengshi Zheng.
\newblock Dbt-net: Dual-branch federative magnitude and phase estimation with
  attention-in-attention transformer for monaural speech enhancement.
\newblock {\em IEEE/ACM Transactions on Audio, Speech, and Language
  Processing}, 30:2629--2644, 2022.

\bibitem{zheng2020improving}
Yibin Zheng, Xinhui Li, Fenglong Xie, and Li~Lu.
\newblock Improving end-to-end speech synthesis with local recurrent neural
  network enhanced transformer.
\newblock In {\em ICASSP 2020-2020 IEEE International Conference on Acoustics,
  Speech and Signal Processing (ICASSP)}, pages 6734--6738. IEEE, 2020.

\bibitem{gu2023mamba}
Albert Gu and Tri Dao.
\newblock Mamba: Linear-time sequence modeling with selective state spaces.
\newblock {\em arXiv preprint arXiv:2312.00752}, 2023.

\bibitem{voelker2019legendre}
Aaron Voelker, Ivana Kaji{\'c}, and Chris Eliasmith.
\newblock Legendre memory units: Continuous-time representation in recurrent
  neural networks.
\newblock {\em Advances in neural information processing systems}, 32, 2019.

\bibitem{chilkuri2021parallelizing}
Narsimha~Reddy Chilkuri and Chris Eliasmith.
\newblock Parallelizing legendre memory unit training.
\newblock In {\em International Conference on Machine Learning}, pages
  1898--1907. PMLR, 2021.

\bibitem{10423155}
Linhui Sun, Shuo Yuan, Aifei Gong, Lei Ye, and Eng~Siong Chng.
\newblock Dual-branch modeling based on state-space model for speech
  enhancement.
\newblock {\em IEEE/ACM Transactions on Audio, Speech, and Language
  Processing}, 32:1457--1467, 2024.

\bibitem{pade1892representation}
Henri Pad{\'e}.
\newblock Sur la repr{\'e}sentation approch{\'e}e d'une fonction par des
  fractions rationnelles.
\newblock In {\em Annales scientifiques de l'Ecole normale sup{\'e}rieure},
  volume~9, pages 3--93, 1892.

\bibitem{maass1997networks}
Wolfgang Maass.
\newblock Networks of spiking neurons: the third generation of neural network
  models.
\newblock {\em Neural networks}, 10(9):1659--1671, 1997.

\bibitem{shrestha2024efficient}
Sumit~Bam Shrestha, Jonathan Timcheck, Paxon Frady, Leobardo Campos-Macias, and
  Mike Davies.
\newblock Efficient video and audio processing with loihi 2.
\newblock In {\em ICASSP 2024-2024 IEEE International Conference on Acoustics,
  Speech and Signal Processing (ICASSP)}, pages 13481--13485. IEEE, 2024.

\bibitem{wang2023spatial}
Yuchen Wang, Kexin Shi, Chengzhuo Lu, Yuguo Liu, Malu Zhang, and Hong Qu.
\newblock Spatial-temporal self-attention for asynchronous spiking neural
  networks.
\newblock In {\em IJCAI}, pages 3085--3093, 2023.

\bibitem{neftci2019surrogate}
Emre~O Neftci, Hesham Mostafa, and Friedemann Zenke.
\newblock Surrogate gradient learning in spiking neural networks: Bringing the
  power of gradient-based optimization to spiking neural networks.
\newblock {\em IEEE Signal Processing Magazine}, 36(6):51--63, 2019.

\bibitem{yu2025beyond}
Ziqiao Yu, Pengfei Sun, and Dan~FM Goodman.
\newblock Beyond rate coding: Surrogate gradients enable spike timing learning
  in spiking neural networks.
\newblock {\em arXiv preprint arXiv:2507.16043}, 2025.

\bibitem{yin2021accurate}
Bojian Yin, Federico Corradi, and Sander~M Boht{\'e}.
\newblock Accurate and efficient time-domain classification with adaptive
  spiking recurrent neural networks.
\newblock {\em Nature Machine Intelligence}, 3(10):905--913, 2021.

\bibitem{shrestha2018slayer}
Sumit~B Shrestha and Garrick Orchard.
\newblock Slayer: Spike layer error reassignment in time.
\newblock {\em Advances in neural information processing systems}, 31, 2018.

\bibitem{zhang2020supervised}
Malu Zhang, Jibin Wu, Ammar Belatreche, Zihan Pan, Xiurui Xie, Yansong Chua,
  Guoqi Li, Hong Qu, and Haizhou Li.
\newblock Supervised learning in spiking neural networks with synaptic
  delay-weight plasticity.
\newblock {\em Neurocomputing}, 409:103--118, 2020.

\bibitem{sun2022axonal}
Pengfei Sun, Longwei Zhu, and Dick Botteldooren.
\newblock Axonal delay as a short-term memory for feed forward deep spiking
  neural networks.
\newblock In {\em ICASSP 2022-2022 IEEE International Conference on Acoustics,
  Speech and Signal Processing (ICASSP)}, pages 8932--8936. IEEE, 2022.

\bibitem{sun2025exploitingheterogeneousdelaysefficient}
Pengfei Sun, Jascha Achterberg, Zhe Su, Dan F.~M. Goodman, and Danyal Akarca.
\newblock Exploiting heterogeneous delays for efficient computation in low-bit
  neural networks, 2025.

\bibitem{perez2021neural}
Nicolas Perez-Nieves, Vincent~CH Leung, Pier~Luigi Dragotti, and Dan~FM
  Goodman.
\newblock Neural heterogeneity promotes robust learning.
\newblock {\em Nature communications}, 12(1):5791, 2021.

\bibitem{10094768}
Pengfei Sun, Ehsan Eqlimi, Yansong Chua, Paul Devos, and Dick Botteldooren.
\newblock Adaptive axonal delays in feedforward spiking neural networks for
  accurate spoken word recognition.
\newblock In {\em ICASSP 2023 - 2023 IEEE International Conference on
  Acoustics, Speech and Signal Processing (ICASSP)}, pages 1--5, 2023.

\bibitem{wu2021tandem}
Jibin Wu, Yansong Chua, Malu Zhang, Guoqi Li, Haizhou Li, and Kay~Chen Tan.
\newblock A tandem learning rule for effective training and rapid inference of
  deep spiking neural networks.
\newblock {\em IEEE Transactions on Neural Networks and Learning Systems},
  34(1):446--460, 2021.

\bibitem{du2024spiking}
Yu~Du, Xu~Liu, and Yansong Chua.
\newblock Spiking structured state space model for monaural speech enhancement.
\newblock In {\em ICASSP 2024-2024 IEEE International Conference on Acoustics,
  Speech and Signal Processing (ICASSP)}, pages 766--770. IEEE, 2024.

\bibitem{hinton2012neural}
Geoffrey Hinton, Nitish Srivastava, and Kevin Swersky.
\newblock Neural networks for machine learning lecture 6a overview of
  mini-batch gradient descent.
\newblock {\em Cited on}, 14(8):2, 2012.

\bibitem{bengio2013estimating}
Yoshua Bengio, Nicholas L{\'e}onard, and Aaron Courville.
\newblock Estimating or propagating gradients through stochastic neurons for
  conditional computation.
\newblock {\em arXiv preprint arXiv:1308.3432}, 2013.

\bibitem{abbott2005model}
Larry~F Abbott and Thomas~B Kepler.
\newblock Model neurons: from hodgkin-huxley to hopfield.
\newblock In {\em Statistical Mechanics of Neural Networks: Proceedings of the
  Xlth Sitges Conference Sitges, Barcelona, Spain, 3--7 June 1990}, pages
  5--18. Springer, 2005.

\bibitem{cramer2020heidelberg}
Benjamin Cramer, Yannik Stradmann, Johannes Schemmel, and Friedemann Zenke.
\newblock The heidelberg spiking data sets for the systematic evaluation of
  spiking neural networks.
\newblock {\em IEEE Transactions on Neural Networks and Learning Systems},
  2020.

\bibitem{tempoaralatten}
Man Yao, Huanhuan Gao, Guangshe Zhao, Dingheng Wang, Yihan Lin, Zhaoxu Yang,
  and Guoqi Li.
\newblock Temporal-wise attention spiking neural networks for event streams
  classification.
\newblock In {\em Proceedings of the IEEE/CVF international conference on
  computer vision}, pages 10221--10230, 2021.

\bibitem{sun2023learnable}
Pengfei Sun, Yansong Chua, Paul Devos, and Dick Botteldooren.
\newblock Learnable axonal delay in spiking neural networks improves spoken
  word recognition.
\newblock {\em Frontiers in Neuroscience}, 17:1275944, 2023.

\bibitem{hammouamrilearning}
Ilyass Hammouamri, Ismail Khalfaoui-Hassani, and Timoth{\'e}e Masquelier.
\newblock Learning delays in spiking neural networks using dilated convolutions
  with learnable spacings.
\newblock In {\em The Twelfth International Conference on Learning
  Representations}.

\bibitem{d2024denram}
Simone D’agostino, Filippo Moro, Tristan Torchet, Yi{\u{g}}it Demira{\u{g}},
  Laurent Grenouillet, Niccol{\`o} Castellani, Giacomo Indiveri, Elisa
  Vianello, and Melika Payvand.
\newblock Denram: neuromorphic dendritic architecture with rram for efficient
  temporal processing with delays.
\newblock {\em Nature communications}, 15(1):3446, 2024.

\bibitem{patino2024hardware}
Alberto Patino-Saucedo, Roy Meijer, Amirreza Yousefzadeh, Manil-Dev Gomony,
  Federico Corradi, Paul Detteter, Laura Garrido-Regife, Bernabe
  Linares-Barranco, and Manolis Sifalakis.
\newblock Hardware-aware training of models with synaptic delays for digital
  event-driven neuromorphic processors.
\newblock {\em arXiv preprint arXiv:2404.10597}, 2024.

\bibitem{malettira2024tskips}
Prajna~G Malettira, Shubham Negi, Wachirawit Ponghiran, and Kaushik Roy.
\newblock Tskips: Efficiency through explicit temporal delay connections in
  spiking neural networks.
\newblock {\em arXiv preprint arXiv:2411.16711}, 2024.

\bibitem{meszaros2025efficient}
Bal{\'a}zs M{\'e}sz{\'a}ros, James~C Knight, and Thomas Nowotny.
\newblock Efficient event-based delay learning in spiking neural networks.
\newblock {\em arXiv preprint arXiv:2501.07331}, 2025.

\bibitem{queant2025delrec}
Alexandre Queant, Ulysse Ran{\c{c}}on, Benoit~R Cottereau, and Timoth{\'e}e
  Masquelier.
\newblock Delrec: learning delays in recurrent spiking neural networks.
\newblock {\em arXiv preprint arXiv:2509.24852}, 2025.

\bibitem{sun2025towards}
Pengfei Sun, Jibin Wu, Paul Devos, and Dick Botteldooren.
\newblock Towards parameter-free attentional spiking neural networks.
\newblock {\em Neural Networks}, 185:107154, 2025.

\bibitem{shrestha2022spikemax}
Sumit~Bam Shrestha, Longwei Zhu, and Pengfei Sun.
\newblock Spikemax: spike-based loss methods for classification.
\newblock In {\em 2022 international joint conference on neural networks
  (IJCNN)}, pages 1--7. IEEE, 2022.

\bibitem{cook2025brain}
Jack Cook, Danyal Akarca, Rui~Ponte Costa, and Jascha Achterberg.
\newblock Brain-like processing pathways form in models with heterogeneous
  experts.
\newblock {\em arXiv preprint arXiv:2506.02813}, 2025.

\bibitem{yin2020effective}
Bojian Yin, Federico Corradi, and Sander~M Boht{\'e}.
\newblock Effective and efficient computation with multiple-timescale spiking
  recurrent neural networks.
\newblock In {\em International Conference on Neuromorphic Systems 2020}, pages
  1--8, 2020.

\bibitem{zhang2024tc}
Shimin Zhang, Qu~Yang, Chenxiang Ma, Jibin Wu, Haizhou Li, and Kay~Chen Tan.
\newblock Tc-lif: A two-compartment spiking neuron model for long-term
  sequential modelling.
\newblock In {\em Proceedings of the AAAI Conference on Artificial
  Intelligence}, volume~38, pages 16838--16847, 2024.

\bibitem{yao2021temporal}
Man Yao, Huanhuan Gao, Guangshe Zhao, Dingheng Wang, Yihan Lin, Zhaoxu Yang,
  and Guoqi Li.
\newblock Temporal-wise attention spiking neural networks for event streams
  classification.
\newblock In {\em Proceedings of the IEEE/CVF International Conference on
  Computer Vision}, pages 10221--10230, 2021.

\bibitem{hamdi2022attention}
Skander Hamdi, Mourad Oussalah, Abdelouahab Moussaoui, and Mohamed Saidi.
\newblock Attention-based hybrid cnn-lstm and spectral data augmentation for
  covid-19 diagnosis from cough sound.
\newblock {\em Journal of Intelligent Information Systems}, 59(2):367--389,
  2022.

\bibitem{orlandic2021coughvid}
Lara Orlandic, Tomas Teijeiro, and David Atienza.
\newblock The coughvid crowdsourcing dataset, a corpus for the study of
  large-scale cough analysis algorithms.
\newblock {\em Scientific Data}, 8(1):156, 2021.

\bibitem{conguh}
Mesut Melek.
\newblock Diagnosis of covid-19 and non-covid-19 patients by classifying only a
  single cough sound.
\newblock {\em Neural Computing and Applications}, 33(24):17621--17632, 2021.

\bibitem{sun2024eeg}
Pengfei Sun, Jorg De~Winne, Paul Devos, and Dick Botteldooren.
\newblock Eeg decoding with conditional identification information.
\newblock {\em Advances in Signal Processing and Artificial Intelligence}, page
  130, 2024.

\bibitem{sun2025electroencephalography}
Pengfei Sun, Jorg De~Winne, Malu Zhang, Paul Devos, and Dick Botteldooren.
\newblock Electroencephalography decoding with conditional identification
  generator.
\newblock {\em International Journal of Neural Systems}, 35(07), 2025.

\bibitem{lawhern2018eegnet}
Vernon~J Lawhern, Amelia~J Solon, Nicholas~R Waytowich, Stephen~M Gordon,
  Chou~P Hung, and Brent~J Lance.
\newblock Eegnet: a compact convolutional neural network for eeg-based
  brain--computer interfaces.
\newblock {\em Journal of neural engineering}, 15(5):056013, 2018.

\bibitem{warden2018speech}
Pete Warden.
\newblock Speech commands: A dataset for limited-vocabulary speech recognition.
\newblock {\em arXiv preprint arXiv:1804.03209}, 2018.

\bibitem{sc1}
Yuan Gong, Yu-An Chung, and James Glass.
\newblock Ast: Audio spectrogram transformer.
\newblock {\em arXiv preprint arXiv:2104.01778}, 2021.

\bibitem{UniRepLKNe}
Xiaohan Ding, Yiyuan Zhang, Yixiao Ge, Sijie Zhao, Lin Song, Xiangyu Yue, and
  Ying Shan.
\newblock Unireplknet: A universal perception large-kernel convnet for audio
  video point cloud time-series and image recognition.
\newblock In {\em Proceedings of the IEEE/CVF conference on computer vision and
  pattern recognition}, pages 5513--5524, 2024.

\bibitem{yilmaz2020deep}
Emre Y{\i}lmaz, Ozg{\"u}r~Bora Gevrek, Jibin Wu, Yuxiang Chen, Xuanbo Meng, and
  Haizhou Li.
\newblock Deep convolutional spiking neural networks for keyword spotting.
\newblock In {\em Proceedings of INTERSPEECH}, pages 2557--2561, 2020.

\bibitem{salaj2021spike}
Darjan Salaj, Anand Subramoney, Ceca Kraisnikovic, Guillaume Bellec, Robert
  Legenstein, and Wolfgang Maass.
\newblock Spike frequency adaptation supports network computations on
  temporally dispersed information.
\newblock {\em Elife}, 10:e65459, 2021.

\bibitem{bittar2022surrogate}
Alexandre Bittar and Philip~N Garner.
\newblock A surrogate gradient spiking baseline for speech command recognition.
\newblock {\em Frontiers in Neuroscience}, 16:865897, 2022.

\bibitem{de2018neural}
Douglas~Coimbra De~Andrade, Sabato Leo, Martin Loesener Da~Silva Viana, and
  Christoph Bernkopf.
\newblock A neural attention model for speech command recognition.
\newblock {\em arXiv preprint arXiv:1808.08929}, 2018.

\bibitem{liulmuformer}
Zeyu Liu, Gourav Datta, Anni Li, and Peter~Anthony Beerel.
\newblock Lmuformer: Low complexity yet powerful spiking model with legendre
  memory units.
\newblock In {\em The Twelfth International Conference on Learning
  Representations}.

\bibitem{chandar2019towards}
Sarath Chandar, Chinnadhurai Sankar, Eugene Vorontsov, Samira~Ebrahimi Kahou,
  and Yoshua Bengio.
\newblock Towards non-saturating recurrent units for modelling long-term
  dependencies.
\newblock In {\em Proceedings of the AAAI Conference on Artificial
  Intelligence}, volume~33, pages 3280--3287, 2019.

\bibitem{erichson2020lipschitz}
N~Benjamin Erichson, Omri Azencot, Alejandro Queiruga, Liam Hodgkinson, and
  Michael~W Mahoney.
\newblock Lipschitz recurrent neural networks.
\newblock In {\em International Conference on Learning Representations}, 2020.

\bibitem{rusch2020coupled}
T~Konstantin Rusch and Siddhartha Mishra.
\newblock Coupled oscillatory recurrent neural network (cornn): An accurate and
  (gradient) stable architecture for learning long time dependencies.
\newblock In {\em International Conference on Learning Representations}, 2020.

\bibitem{le2015simple}
Quoc~V Le, Navdeep Jaitly, and Geoffrey~E Hinton.
\newblock A simple way to initialize recurrent networks of rectified linear
  units.
\newblock {\em arXiv preprint arXiv:1504.00941}, 2015.

\bibitem{rusch2021long}
T~Konstantin Rusch, Siddhartha Mishra, N~Benjamin Erichson, and Michael~W
  Mahoney.
\newblock Long expressive memory for sequence modeling.
\newblock {\em arXiv preprint arXiv:2110.04744}, 2021.

\bibitem{psmnist}
David~W Romero, David~M Knigge, Albert Gu, Erik~J Bekkers, Efstratios Gavves,
  Jakub~M Tomczak, and Mark Hoogendoorn.
\newblock Towards a general purpose cnn for long range dependencies in $ n $ d.
\newblock {\em arXiv preprint arXiv:2206.03398}, 2022.

\end{thebibliography}
\vfill
\end{document}